\documentclass[twocolumn,showpacs,preprintnumbers,amsmath,amssymb]{revtex4}

\usepackage{graphicx}
\usepackage{dcolumn}
\usepackage{bm}

\begin{document}

\title{Theory of single photon detection by 'dirty' current-carrying superconducting
strip based on the kinetic equation approach}

\author{D.Yu. Vodolazov}
\affiliation{Institute for Physics of Microstructure, Russian
Academy of Sciences, 603950, Nizhny Novgorod, GSP-105, Russia}

\date{\today}


\begin{abstract}

Using kinetic equation approach we study dynamics of electrons and
phonons in current-carrying superconducting nanostrips after
absorption of single photon of near-infrared or optical range. We
find that the larger the ratio $C_e/C_{ph}|_{T_c}$ ($T_c$ is a
critical temperature of superconductor, $C_e$ and $C_{ph}$ are
specific heat capacities of electrons and phonons, respectively)
the larger part of photon's energy goes to electrons, they become
stronger heated and, hence, could thermalize faster during initial
stage of hot spot formation. Thermalization time $\tau_{th}$ could
be less then one picosecond for superconductors with
$C_e/C_{ph}|_{T_c}\gg 1$ and small diffusion coefficient $D\simeq
0.5 cm^2/s$ when thermalization occurs mainly due to
electron-phonon and phonon-electron scattering in relatively small
volume $\sim \xi^2d$ ($\xi$ is a superconducting coherence length,
$d<\xi$ is a thickness of the strip). At larger times because of
diffusion of hot electrons effective temperature inside the hot
spot decreases, the size of hot spot increases, superconducting
state becomes unstable and normal domain spreads in the strip at
current larger than so-called detection current. We find
dependence of detection current on the photon's energy, place of
its absorption in the strip, width of the strip, magnetic field
and compare it with existing experiments. Our results demonstrate
that materials with $C_e/C_{ph}|_{T_c} \ll 1$ are bad candidates
for single photon detectors due to small transfer of photon's
energy to electronic system and large $\tau_{th}$. We also predict
that even several microns wide dirty superconducting bridge is
able to detect single near-infrared or optical photon if its
critical current exceeds 70 $\%$ of depairing current and
$C_e/C_{ph}|_{T_c} \gtrsim 1$.

\end{abstract}

\maketitle

\section{Introduction}

Main idea of single photon detection by the current carrying
superconducting strip is relatively simple. The absorbed photon
heats the electrons in the restricted area of the strip (it is
called as a hot spot), superconductivity is locally destroyed and
critical current $I_c$ of the strip is reduced up to
$I_c^{spot}<I_c$. If transport current exceeds $I_c^{spot}$
transition to the resistive state occurs and it is detected in the
experiment.

At the present time there are several phenomenological models
which offer different scenarios for appearance of the resistive
response after photon absorption and which explain some
experimental results
\cite{Semenov_EPJB,Vodolazov_PRB_1,Bulaevskii_PRB,Eftekharian,Zotova_SUST,Engel_IEEE_model}
(see also recent review of models in Ref.
\cite{Engel_SUST_review}). The drawback of these models is that
they use phenomenological assumptions about size of the hot
region, level of suppression of superconductivity and part of the
photon's energy which is stored in the electronic system. Besides,
some of these models
\cite{Semenov_EPJB,Eftekharian,Engel_IEEE_model} operate with the
number of nonequilibrium quasiparticles in order to find the
suppression of magnitude of superconducting order parameter
$|\Delta|$. One should have in mind, that this approach is
quantitatively valid only when the deviation of quaisparticle
(electron) distribution function $n(\epsilon)$ from equilibrium
occurs in the narrow energy interval $\delta \epsilon$ near the
superconducting gap $\epsilon_g \gg \delta \epsilon$ which is not
true in the case of the hot spot formation in thin superconducting
strip, especially at its initial stage, when the effective
temperature of electrons could be larger than the bath temperature
in several times. Therefore one can expect only qualitative
validity of the approaches using ideas of Rothwarf-Taylor model
\cite{RT}.

Below we formulate two problems which we solve to understand what
material could be the good candidate for usage in superconducting
nanowire single photon detector. First problem concerns the
question what part of the energy of the photon goes to the
electronic system and how it is related with material parameters
of the superconducting strip. The second problem is connected with
the question how fast the electrons (throughout the paper by
electrons we mean quasiparticle excitations) are thermalized and
what is the role of electron-electron inelastic scattering. It is
known, that all materials which show good ability to detect single
photons are extremely dirty superconducting strips with small
diffusion coefficient $D\simeq 0.5 cm^2/s$ and low critical
temperature $T_c \lesssim 10 K$. First of all, small $D$ does not
allow fast diffusion of electrons which favors their fast
thermalization, because the energy of absorbed photon is confined
in relatively small volume at initial stage of hot spot formation
leading to relatively high 'temperature' of hot electrons.
Secondly, the smaller $D$ the smaller is the electron-electron
inelastic relaxation time $\tau_{e-e}$ which also decreases
thermalization time $\tau_{th}$ and increases the capability to
detect single photon. Indeed, if thermalization time $\tau_{th}$
of electrons is larger than their diffusion time $\tau_{D,w}\simeq
w^2/4D$ across the strip with width $w$ the photon's energy will
be smeared over the large area which leads to smaller influence on
superconducting properties and it complicates photon's detection.

To answer above questions we numerically solve kinetic equations
for electron and phonon distribution functions, taking into
account electron-phonon, phonon-electron and electron-electron
scattering. First of all we study the initial stage of the
electron-phonon downconversion cascade on time scale comparable
with the characteristic time of variation of $|\Delta|$ -
$\tau_{\Delta}\simeq \hbar/|\Delta| \simeq \hbar/k_BT_c$ during
which one cannot expect strong suppression of superconducting
order parameter and electrons diffuse only on distance $\sim
\sqrt{D\tau_{|\Delta|}}$ which is about of superconducting
coherence length in dirty superconductors $\xi\sim \sqrt{\hbar
D/|\Delta|}$. We find that electron-phonon downconversion cascade
and thermalization in electronic system depends not only on
strength of electron-electron scattering, but also on ratio of
electronic $C_e$ and phonon $C_{ph}$ heat capacities taken at
$T=T_c$. Indeed, the larger the ratio $C_e/C_{ph}|_{T_c}$ the
larger part of phonon's energy goes to electronic system, its
effective 'temperature' becomes higher and thermalization time
$\tau_{th}$ is shorter. We show that relatively short
$\tau_{th}\simeq \tau_{|\Delta|}$ could be reached in
superconductors with large ratio $C_e/C_{ph}|_{T_c} \gg 1$ when
thermalization occurs mainly via electron-phonon and
phonon-electron scattering.

Dynamics of hot spot at times $t\gtrsim \tau_{|\Delta|}$ we study
in two limits. In the limit of short thermalization time
($\tau_{th}\simeq \tau_{|\Delta|}\ll \tau_{D,w}$) we use two
temperature model and solve heat conductance equations (taking
into account the Joule dissipation) for electron and phonon
temperatures coupled with modified Ginzburg-Landau equation for
superconducting order parameter. In the limit of large
thermalization time ($\tau_{th} \simeq \tau_{D,w}$) we assume
uniform distribution of electron and phonon effective temperatures
across the strip by time $t \simeq \tau_{D,w}$. In both limits we
find the current-energy relation, dependence of the cut-off
photon's energy on the temperature at fixed ratio $I/I_{dep}(T)$
($I_{dep}$ is a depairing current), study the role of the magnetic
field and how current-energy relation depends on the strip's
width. We show, that relatively narrow (with width about 100-200
nm) superconducting strip with small ratio $C_e/C_{ph}|_{T_c} \ll
1$ needs current close to $I_{dep}$ to be able to detect single
photon with energy about 1 eV. On contrary, such a strip with
ratio $C_e/C_{ph}|_{T_c} \gtrsim 1$ can detect the same photon at
the current much smaller than depairing current. And, finally, we
predict that even wide superconducting strip with
$C_e/C_{ph}|_{T_c} \gtrsim 1$ and width of about several microns
can detect single infrared or optical photon if the strip can be
biased, without a loss of superconductivity, at $I \gtrsim 0.7
I_{dep}(T)$.

The structure of the paper is following. In Sec. II we present
basic equations. In Sec. III we show results on the initial stage
of hot spot formation (on time scale about of $\tau_{|\Delta|}$
after photon's absorption) when its radius is smaller than the
coherence length. In Sec. IV we study the case when instability of
superconducting state occurs at the moment when the hot electrons
reach both edges of the strip and form hot belt (limit of large
$\tau_{th}$), while in Sec. V we consider the opposite case when
superconducting state becomes unstable before the hot spot expands
across the strip (limit of small $\tau_{th}$). In Sec. VI we
compare our results with existing theories and experiments. In
Sec. VII we formulate our main results.

\section{Equations}

In this section we present equations which we use to study the
dynamics of electron and phonon distribution functions in
superconducting strip at the initial stage of hot spot formation
after absorption of the single photon. First of all these are the
kinetic equations for electron $n_{\epsilon}$ and phonon
$N_{\epsilon}$ distribution functions

\begin{widetext}
\begin{eqnarray}
N_1\frac{\partial n}{\partial t }= D\nabla((N_1^2-R_2^2)\nabla n)-
R_2\frac{\partial n}{\partial \epsilon}\frac{\partial
|\Delta|}{\partial t}+ I_{e-ph}(n,N)+I_{e-e}(n),
\\
\frac{\partial N}{\partial t} = -\frac{N-N_{eq}(T)}{\tau_{esc}}+
I_{ph-e}(N,n),
\end{eqnarray}
where $I_{e-ph}(n,N)$, $I_{ph-e}(N,n)$ and  $I_{e-e}(n)$ are the
electron-phonon, phonon-electron and electron-electron collision
integrals

\begin{eqnarray}
I_{e-ph}(n,N)=-\frac{1}{(k_BT_c)^3}\frac{1}{\tau_0} [\nonumber
\\
\int_{0}^{\epsilon}
d\epsilon_1M(\epsilon,\epsilon_1)(\epsilon-\epsilon_1)^2
\left(\left[1+2N_{\epsilon-\epsilon_1}\right](n_{\epsilon}-n_{\epsilon_1})+n_{\epsilon}(1-2n_{\epsilon_1})+n_{\epsilon_1}\right)+
\nonumber
\\
\int_{\epsilon}^{\epsilon+\hbar \omega_D}
d\epsilon_1M(\epsilon,\epsilon_1)(\epsilon-\epsilon_1)^2
\left(\left[1+2N_{\epsilon_1-\epsilon}\right](n_{\epsilon}-n_{\epsilon_1})-n_{\epsilon}(1-2n_{\epsilon_1})-n_{\epsilon_1}\right)+
\nonumber
\\
\int_{0}^{\hbar \omega_D-\epsilon}
d\epsilon_1M(\epsilon,-\epsilon_1)(\epsilon+\epsilon_1)^2
\left(\left[1+2N_{\epsilon_1+\epsilon}\right](n_{\epsilon}+n_{\epsilon_1}-1)-n_{\epsilon}(1-2n_{\epsilon_1})-n_{\epsilon_1}+1\right)],
\end{eqnarray}

\begin{eqnarray}
I_{ph-e}(N,n)=\frac{\gamma}{\tau_0k_BT_c} [\int_{0}^{\epsilon}
d\epsilon_1 M(\epsilon_1,-(\epsilon-\epsilon_1))\times \nonumber
\\
\times\left(n_{\epsilon_1}n_{\epsilon-\epsilon_1}+N_{\epsilon}(n_{\epsilon-\epsilon_1}+n_{\epsilon_1}-1)\right)+
\nonumber
\\
+2\int_{0}^{\infty} d\epsilon_1
M(\epsilon_1,(\epsilon+\epsilon_1))
\left((1-n_{\epsilon_1})n_{\epsilon+\epsilon_1}+N_{\epsilon}(n_{\epsilon+\epsilon_1}-n_{\epsilon_1})\right)],
\end{eqnarray}

$$
M(\epsilon,\pm \epsilon_1)=N_1(\epsilon)N_1(\epsilon_1)\mp
R_2(\epsilon)R_2(\epsilon_1),
$$

\begin{eqnarray}
I_{e-e}(n)=\frac{\alpha_{e-e}}{\tau_0k_BT_c}\int_{0}^{\infty}\int_{0}^{\infty}d\epsilon_1d\epsilon_2
[ \nonumber
\\
+\frac{E_1}{|\epsilon-\epsilon_1|}(n_{\epsilon}(1-n_{\epsilon_1})(1-n_{\epsilon_2})(1-n_{\epsilon-\epsilon_1-\epsilon_2})
-(1-n_{\epsilon})n_{\epsilon_1}n_{\epsilon_2}n_{\epsilon-\epsilon_1-\epsilon_2})H_v(\epsilon-\epsilon_1-\epsilon_2)
\nonumber
\\
+E_2\left(\frac{1}{|\epsilon+\epsilon_1|}+\frac{2}{|\epsilon-\epsilon_2|}\right)(n_{\epsilon}n_{\epsilon_1}(1-n_{\epsilon_2})(1-n_{\epsilon+\epsilon_1-\epsilon_2})
-(1-n_{\epsilon})(1-n_{\epsilon_1})n_{\epsilon_2}n_{\epsilon+\epsilon_1-\epsilon_2})H_v(\epsilon+\epsilon_1-\epsilon_2)
\nonumber
\\
+E_3\left(\frac{1}{|\epsilon-\epsilon_1|}+\frac{2}{|\epsilon+\epsilon_2|}\right)(n_{\epsilon}(1-n_{\epsilon_1})n_{\epsilon_2}n_{\epsilon_1-\epsilon_2-\epsilon}
-(1-n_{\epsilon})n_{\epsilon_1}(1-n_{\epsilon_2})(1-n_{\epsilon_1-\epsilon_2-\epsilon})H_v(\epsilon_1-\epsilon_2-\epsilon)],
\end{eqnarray}

\begin{eqnarray}
E_1=a(N_1(\epsilon)N_1(\epsilon_1)N_1(\epsilon_2)N_1(\epsilon-\epsilon_1-\epsilon_2)
-R_2(\epsilon)R_2(\epsilon_1)R_2(\epsilon_2)R_2(\epsilon-\epsilon_1-\epsilon_2))
\nonumber
\\
+b(N_1(\epsilon)R_2(\epsilon_1)R_2(\epsilon_2)N_1(\epsilon-\epsilon_1-\epsilon_2)-
R_2(\epsilon)N_1(\epsilon_1)N_1(\epsilon_2)R_2(\epsilon-\epsilon_1-\epsilon_2)).
\nonumber
\end{eqnarray}
\end{widetext}

Coefficients $E_2$ and $E_3$ are expressed via $E_1$ in the
following way: $E_2=E_1(\epsilon_1\to -\epsilon_1)$,
$E_3=E_1(\epsilon\to -\epsilon, \epsilon_1\to -\epsilon_1)$. $a$
and $b$ are coefficients of order of unity
\cite{Eliashberg,Gulyan} and $H_v(x)$ is a Heaviside function.

$I_{e-ph}(n,N)$ and $I_{ph-e}(N,n)$ are written above for case
when one can neglect renormalization of electron-phonon coupling
constant due to disorder \cite{Kaplan,Chang} and
\begin{equation}
\frac{1}{\tau_0}=g\left(\frac{k_BT_c}{\hbar
\omega_D}\right)^2\frac{k_BT_c}{\hbar},
\end{equation}
is the familiar characteristic time introduced in Ref.
\cite{Kaplan} ($\omega_D$ is a Debye frequency and $g$ is an
electron-phonon coupling constant). Coefficient
\begin{equation}
\gamma=\frac{4\hbar \omega_D N(0)}{9N_{ion}}\left(\frac{\hbar
\omega_D }{k_BT_c}\right)^2=
\frac{8\pi^2}{5}\frac{C_{e}}{C_{ph}}\biggl|_{T=T_c} ,
\end{equation}
staying in front of $I_{ph-e}$ collision integral is proportional
to the ratio of the electronic $C_e(T)=(2\pi^2/3)k_B^2N(0)T$ and
phonon $C_{ph}(T)=(12\pi^4/5)N_{ion}k_B(k_BT/\hbar \omega_D)^3$
specific heat capacities at $T=T_c$ ($N(0)$ is the one spin
density of states of electrons in the normal state at the Fermi
energy $E_F$, $N_{ion}$ is the density of ions).

Electron-electron collision integral in form of Eq. (5) is written
for dirty quasi-2D metallic superconducting film with renormalized
electron-electron coupling constant due to impurities. Coefficient
$\alpha_{e-e}$
\begin{equation}
\alpha_{e-e}=\tau_0\frac{k_BT_c}{4\hbar}\frac{R_{\Box}}{R_Q},
\end{equation}
describes the strength of electron-electron inelastic scattering
($R_{\Box}$ is the sheet resistance and $R_Q=2\pi\hbar/e^2\simeq
25.8 \, kOhm$ is the quantum resistance). In pure superconducting
metal coefficients $1/|\epsilon \pm \epsilon_{1,2}|$ are absent
and Eq. (5) transfers (with $\alpha_{e-e}=\tau_0
(k_BT_c)^2/(2\hbar E_F)$) to expression present in Refs.
\cite{Eliashberg,Gulyan}. In the normal state $E_1=E_2=E_3=a$ and
Eq. (5) coincides (with $\alpha_{e-e}=\tau_0k_BT_c/(4\hbar k_F\l)$
and $a=1$) with e-e collision integral for 2D dirty metal from
Ref. \cite{Rammer}. For dirty quasi-2D metallic normal film from
Eqs. (1,5) and $\epsilon \gg k_BT$ it follows familiar inelastic
e-e scattering time $\tau_{e-e}(\epsilon)=4\hbar R_Q/(\epsilon
R_{\Box})$ (see Eq. (4.4) in \cite{Altshuler_book}).

Because coefficients $a$ and $b$ are unknown for any metal we put
$a=1$ and $b=0$. Choice of $b=0$ is connected with different
expressions for $E_i$ behind $b$ which are present in Refs.
\cite{Eliashberg,Gulyan}. If we choose expression for $E_i$ from
Ref. \cite{Gulyan} (as we do in our work) finite $b>0$ leads to
increase of $\tau_{e-e}$ for electrons having energy close to
$|\Delta|$.

Spectral functions $N_1(\epsilon)$, $R_2(\epsilon)$, entering Eqs.
(1, 3-5), has to be found in the dirty limit from the Usadel
equation
\begin{equation}
\hbar D \nabla^2 \Theta+ \left(2i\epsilon-\frac{D}{\hbar} q_s^2
\cos\Theta\right)\sin\Theta+2|\Delta|\cos\Theta=0,
\end{equation}
where $q_s=mv_s=\hbar(\nabla \phi-2eA/\hbar c)$ is the superfluid
momentum, $\phi$ is a phase of superconducting order parameter
$\Delta=|\Delta|e^{i\phi}$,
$\cos\Theta=N_1(\epsilon)+iR_1(\epsilon)$ and
$\sin\Theta=N_2(\epsilon)+iR_2(\epsilon)$. $N_1(\epsilon)N(0)$ has
a meaning of density of states of electrons in the superconducting
state, while $R_2$ enters the equation for the superconducting
order parameter.

A static self-consistency equation for the magnitude of the order
parameter has a following form
\begin{eqnarray}
\frac{1}{\lambda_{BCS}}=\int_{0}^{\hbar \omega_D}
\frac{R_2}{|\Delta|}(1-2n_{\epsilon})d\epsilon= \nonumber
\\
\int_{0}^{\hbar \omega_D}
\frac{R_2}{|\Delta|}(1-2n_{\epsilon}^{eq})d\epsilon-\Phi_{neq},
\end{eqnarray}
where $n_{\epsilon}^{eq}=1/(exp(\epsilon/k_BT)+1)$ is an
equilibrium distribution function of electrons (quasiparticles)
and $\lambda_{BSC}$ is a coupling constant in BCS theory.
Suppression of $|\Delta|$ due to hot electrons is described by
$\Phi_{neq}$
\begin{equation}
\Phi_{neq}=2\int_{0}^{\hbar \omega_D}
\frac{R_2}{|\Delta|}(n_{\epsilon}-n_{\epsilon}^{eq})d\epsilon.
\end{equation}

Very often, to describe suppression of the order parameter due to
$n_{\epsilon} \neq n_{\epsilon}^{eq}$ the coordinate-dependent
density of nonequilibrium electrons is used, which is determined
as
\begin{eqnarray}
N_{neq}(\vec{r})/V=4N(0)\int_0^{\infty}N_1(n_{\epsilon}(\vec{r})-n_{\epsilon}^{eq})d\epsilon=
\nonumber
\\
 4N(0)\int_{|\Delta|}^{\infty}
\frac{\epsilon(n_{\epsilon}(\vec{r})-n_{\epsilon}^{eq})}{\sqrt{\epsilon^2-|\Delta|^2}}d\epsilon.
\end{eqnarray}

The last expression is valid when one can neglect the gradient
term and term with $q_s^2$ in Usadel equation which leads to
\begin{equation}
N_1(\epsilon)=\frac{\epsilon}{\sqrt{\epsilon^2-|\Delta|^2}}H_v(\epsilon-|\Delta|),
\end{equation}
and
\begin{equation}
R_2(\epsilon)=\frac{|\Delta|}{\sqrt{\epsilon^2-|\Delta|^2}}H_v(\epsilon-|\Delta|).
\end{equation}

Potential $\Phi_{neq}$ could be expressed via $N_{neq}/V$ when
deviation from equilibrium occurs in narrow energy interval near
$|\Delta|$ and one can replace $\epsilon \simeq |\Delta|$ in
numerator of Eq. (12) and take it off the integrand
\begin{equation}
\Phi_{neq}(\vec{r})=2\int_{|\Delta|}^{\hbar \omega_D}
\frac{n(\epsilon,\vec{r})-n_{eq}(\epsilon)}{\sqrt{\epsilon^2-|\Delta|^2}}d\epsilon
\simeq \frac{N_{neq}(\vec{r})}{2N(0)|\Delta|V}.
\end{equation}

When the deviation from equilibrium occurs in wide energy interval
and/or it occurs at energies $\epsilon \gg |\Delta|$ then
$\Phi_{neq} \neq N_{neq}/2N(0)|\Delta|V$ due to presence of extra
factor $\epsilon$ in numerator of Eq. (12). In this case usage of
the approach with number of nonequilibrium electrons cannot
pretend for quantitative description and may be used only for
qualitative analysis.

\section{Initial stage of hot spot formation}

When deviation $n_{\epsilon}$ from $n_{\epsilon}^{eq}$ occurs in
the volume smaller than $\xi^3$ and on time scale shorter than
variation time of $|\Delta|$ one cannot expect strong suppression
of superconductivity. Therefore, as a first approximation, at
times $t<\tau_{|\Delta|}$ we study dynamics of $n_{\epsilon}$ and
$N_{\epsilon}$ after absorption of the photon with
$|\Delta|=const$. To simplify further the problem we also assume
that the energy of the photon is distributed instantly over the
volume $V_{init}=\pi \xi^2 d$ (where $d<\xi$ is a thickness of the
strip). In reality it takes time $\sim \xi^2/D=\tau_{|\Delta|}$
and below we argue that such a simplification should not change
the main result of this section.

With above assumptions we numerically solve Eqs. (1-5), where we
omit gradient terms and consider the case of zero current $I=0$
(finite $I$ leads to smearing of spectral functions at $\epsilon
\simeq |\Delta|$ and does not influence our main result). We use
different initial conditions, which corresponds to different
physical situations. Electronic bubble initial condition
\begin{eqnarray} 
n_{\epsilon}(t=0)=n_{\epsilon}^{eq}+ \frac{\alpha
e^{-(\epsilon-\epsilon_0)^2/\delta \epsilon^2}}{\sqrt{\pi} \delta
\epsilon}, \nonumber
\\
N_{\epsilon}(t=0)=N_{\epsilon}^{eq},
\end{eqnarray}
($N_{\epsilon}^{eq}=1/(exp(\epsilon/k_BT)-1)$ is an equilibrium
distribution function of phonons) corresponds to absorption of the
photon and creation of initial hot electrons at energy $\epsilon
\simeq \epsilon_0 \gg \delta \epsilon$.

Phonon bubble initial condition
\begin{eqnarray}
n_{\epsilon}(t=0)=n_{\epsilon}^{eq},
 \nonumber
\\
N_{\epsilon}(t=0)=N_{\epsilon}^{eq}+ \frac{\beta
e^{-(\epsilon-\epsilon_0)^2/\delta \epsilon^2}}{\sqrt{\pi} \delta
\epsilon},
\end{eqnarray}
models situation when, for example, the molecule hits the strip
and excites phonons with energy $\epsilon \simeq \epsilon_0$.
Instead of photon bubble one can use phonon plateau initial
condition (when acoustic phonons of all available energies $0 \leq
\epsilon \leq \hbar\omega_D$ are excited with equal probability)
and results are practically undistinguishable from the phonon
bubble initial condition (if one is interested in thermalization
time and dynamics of energy contained in electronic and phonon
systems).

Third initial condition corresponds to extreme case of very high
e-e relaxation rate, which at all energies $\epsilon < E_{photon}$
exceeds e-ph relaxation rate and at $\epsilon \simeq k_B T$ is
larger than $1/\tau_{|\Delta|}$. In this situation electrons are
thermalized at times $t\ll \tau_{|\Delta|}$ and all energy of the
photon is kept in electronic system at $t=0$
\begin{eqnarray}
n_{\epsilon}(t=0)=\frac{1}{e^{\epsilon/k_BT_e}+1} \nonumber,
\\
N_{\epsilon}(t=0)=N_{\epsilon}^{eq}.
\end{eqnarray}

In all cases we choose parameters $\alpha$, $\beta$ and $T_e$ in
Eqs. (16-18) in a way to keep absorbed energy per unit of volume
the same. For electron bubble condition we choose $\epsilon_0 \gg
\hbar\omega_D$, while for phonon bubble condition $\epsilon_0
\simeq \hbar\omega_D$.

During calculations we check that the energy is conserved
\begin{equation}
E_{photon}/V_{init}=(E_{ph}+E_e)/V_{init}-(E_{ph}+E_e)^{eq}/V_{init},
\end{equation}
where $E_{ph}$ is the energy of phonon system in the Debye model
with quadratic density of states
$\mathcal{D}(\epsilon)=9\epsilon^2/\hbar \omega_D$ per ion
\begin{eqnarray}
E_{ph}/V_{init}=\frac{1}{V_{init}}\frac{\mathcal{D}(\hbar
\omega_D)}{\hbar(\hbar\omega_D)^2}\int_0^{\hbar
\omega_D}\epsilon^3 N_\epsilon d\epsilon= \nonumber
\\
\frac{E_0}{\gamma}\int_{0}^{\hbar\omega_D/k_BT_c}
\tilde{\epsilon}^3N_{\tilde{\epsilon}}d\tilde{\epsilon},
\end{eqnarray}
 $E_e$ is the energy of electrons(quasiparticles) in
the superconductor
\begin{eqnarray}
E_e/V_{init}=
\nonumber
\\
4N(0)\left(\int_0^{\infty}\epsilon N_1 n_{\epsilon}d\epsilon
-\frac{|\Delta|^2}{4}\left(\frac{1}{2}+
ln\left(\frac{\Delta_0}{|\Delta|} \right) \right) \right)=
\nonumber
\\
E_0\left(\int_0^{\infty}N_1\tilde{\epsilon}n_{\tilde{\epsilon}}d\tilde{\epsilon}
-\left(\frac{|\Delta|}{2k_BT_c}\right)^2\left(\frac{1}{2}+
ln\left(\frac{\Delta_0}{|\Delta|}\right)\right)\right),
\end{eqnarray}

with $\tilde{\epsilon}=\epsilon/k_BT_c$, $\Delta_0=1.76 k_BT_c$,
$E_0=4N(0)(k_BT_c)^2$.

In numerical calculations we use parameters typical for NbN:
$\hbar \omega_D=30 meV$ (chosen value of $\hbar \omega_D$ is some
kind of compromise between variety values known for different
phases of NbN \cite{SR_Zou}), $k_BT_c=0.86 meV$ ($T_c=10K$),
$T=T_c/2$, $E_{photon}/E_0V_{init}=60$ (with $N(0)=25.5$
eV$^{-1}$nm$^{-3}$ \cite{Engel_APL}, $\xi=4.8 nm$, $d=4 nm$ it
corresponds to $E_{photon}\simeq 1.3 eV$) and neglect escape of
nonequilibrium phonons to substrate because usually $\tau_{esc}
\gg \tau_{|\Delta|}$.

At time $t \lesssim \tau_{|\Delta|}$ energy of the absorbed photon
$\sim 1.3 eV$ is concentrated in relatively small volume
$V_{init}\simeq \pi\xi^2d \pi \simeq 290 nm^3$ which means high
energy concentration. The larger the $\gamma$ the higher is the
temperature of both electrons $T_e$ and phonons $T_{ph}=T_e$, as
one can see from Eqs. (19-21) if one inserts there Fermi-Dirac and
Bose-Einstein functions for $n_{\epsilon}$ and $N_{\epsilon}$,
respectively (these functions null collision integrals when the
downconversion cascade is over and one neglects diffusion of
electrons). Because electron-phonon relaxation time $\tau_{e-ph}
(\epsilon \ll k_BT_e)\simeq T_e^{-3}$ and $\tau_{e-e}(\epsilon \ll
k_BT_e)\simeq \alpha_{e-e}^{-1}T_e{^{-1}}$ one could expect that
for relatively large $\gamma$ and $\alpha_{e-e}$ the energy of
optical or near-infrared photon could be shared between electron
and phonon systems and electrons with phonons could be thermalized
by time $t\simeq \tau_{|\Delta|}$.
\begin{figure}[hbtp]
\includegraphics[width=0.48\textwidth]{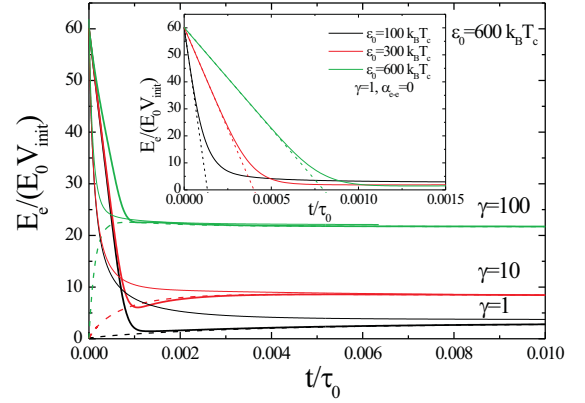}
\caption{Dependence of the electronic energy $E_e$ on time found
from solution of kinetic equations at different $\gamma$ and
different initial conditions: electron bubble with $\epsilon_0=600
k_BT_c$, $\alpha_{e-e}=0$ (solid thick curves), phonon bubble with
$\epsilon_0=30 k_BT_c$, $\alpha_{e-e}=0$ (dashed curves) and hot
electrons with initial $T_e=8.6T_c$, $\alpha_{e-e}=1000$ (solid
thin curves). In all cases the injected energy to the electrons
and phonons is the same and it is equal to $\simeq 1.3 eV$ in the
volume $V_{init}=\pi \xi^2 d \simeq 290$ nm$^3$. In the inset we
show dynamics of $E_e$ with electron bubble initial condition,
$\alpha_{e-e}=0$ and different $\epsilon_0$. Dashed lines are
linear dependence $E_{e}=E_{photon}(1-t/\tau_{leak})$ with
$\tau_{leak}$ taken from Eq. (22).}
\end{figure}

To prove it in Fig. 1 we show the time dependence of the energy of
the electronic system calculated at different initial conditions,
$\gamma=1-100$ and the same injected energy. One can see that for
$\gamma=100$ already by time $\tau_{sh} \simeq 0.001\tau_0$ the
largest part of injected energy is shared between electrons and
phonons (even in absence of e-e scattering) and $\tau_{sh}$
increases with decrease of $\gamma$. Parameter $\gamma$ controls
what part of injected energy goes to electronic system - the
larger $\gamma$ the larger is this part (see Fig. 1).
\begin{figure}[hbtp]
\includegraphics[width=0.53\textwidth]{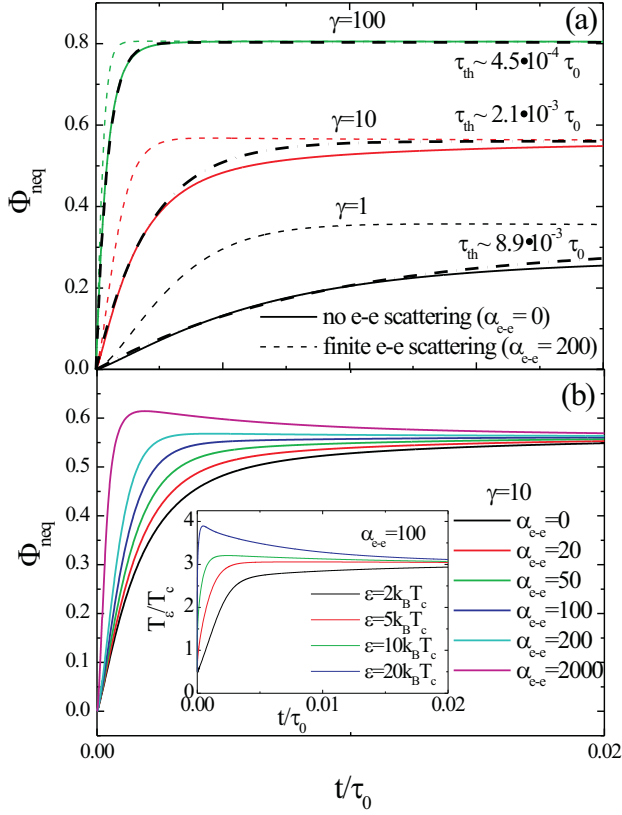}
\caption{(a) Time evolution of $\Phi_{neq}$ at different $\gamma$
and two values of $\alpha_{e-e}$. Fitting $\Phi_{neq}(t)$ (when
$\alpha_{e-e}=0$) by expression
$\Phi_{neq}(t\to\infty)(1-exp(-t/{\tau_{th}}))$ is shown by thick
dashed curves with corresponding $\tau_{th}$. (b) Time evolution
of $\Phi_{neq}$ at $\gamma=10$ and different $\alpha_{e-e}$. All
results are obtained for phonon bubble initial condition and
$T=T_c/2$. In inset of figure (b) we present time evolution of
$T_{\epsilon}=\epsilon/k_Bln(1/n_{\epsilon}-1)$ at different
energies and $\gamma=10$, $\alpha_{e-e}=100$. At t=0,
$n_{\epsilon}=n_{\epsilon}^{eq}$ and $T_{\epsilon}=T=T_c/2$ at all
energies. After injection of energy to the phonon system
$n_{\epsilon}$ deviates from equilibrium which leads to different
$T_{\epsilon}$. $T_{\epsilon}=T_e\simeq 3T_c$ at all energies at
$t\gg \tau_{th}\simeq 1.2 \cdot 10^{-3} \tau_0$.}
\end{figure}

Sharing time $\tau_{sh}$ is actually the thermalization time
$\tau_{th}$ in electronic and phonon systems. In Fig. 2 we show
time evolution of $\Phi_{neq}$ (which controls the suppression of
$|\Delta|$) at different $\gamma$ and $\alpha_{e-e}$. Results are
found in case of phonon bubble initial condition. $\Phi_{neq}$
practically stops depending on time when electrons are thermalized
and $n_{\epsilon}$ is described by Fermi-Dirac function. We fit
numerical $\Phi_{neq}(t)$ by the expression
$\Phi_{neq}(t\to\infty)(1-exp(-t/{\tau_{th}}))$ and find
$\tau_{th}$ (examples of fitting are shown in Fig. 2). One can see
that thermalization time decreases with increase of $\gamma$
(which is consequence of larger energy transfer to electron
system) or with increase of $\alpha_{e-e}$ (which is consequence
of shorter $\tau_{e-e}$).

In case of electron bubble initial condition and $\epsilon_0 \gg
\hbar \omega_D, \delta \epsilon \ll \epsilon_0$ one can find
analytical expression for time $\tau_{leak}$ during which the
energy leaks from electrons to phonons at very beginning of
downconversion cascade. Indeed, inserting $n_{\epsilon}$ in form
of Eq. (16) in phonon-electron collision integral one finds that
$I_{ph-e}\simeq 2 \gamma \alpha /(\tau_0 k_BT_c)$ at energies
$\epsilon \gg |\Delta|$. Than assuming that $\alpha$ depends on
time and the full energy is conserved and is equal to the energy
of the photon (we neglect here $E_e^{eq}+E_{ph}^{eq} \ll
E_{photon}$) one finds $E_e=E_{photon}exp(-t/\tau_{leak})$ with
\begin{equation}
\tau_{leak}=\tau_0\frac{2\epsilon_0 (k_BT_c)^3}{(\hbar
\omega_D)^4}=\frac{2\epsilon_0\hbar}{g(\hbar \omega_D)^2}.
\end{equation}

At $t\ll \tau_{leak}$ one has linear decay
$E_e=E_{photon}(1-t/\tau_{leak})$ and it is shown as dashed lines
in inset of Fig. 1. For $\gamma=100$ $\tau_{leak} \simeq
\tau_{th}$ and above simple calculations are valid only
qualitatively. For $\gamma=1,10$ $\tau_{leak} \ll \tau_{th}$ and
there is good quantitative agreement between numerical and
analytical results at $t \lesssim \tau_{leak}$ (see inset in Fig.
1). When $\tau_{leak} \ll \tau_{th}$ the phonon system absorbs
more energy by time $t\simeq \tau_{leak}$ than it should have at
$t\gg \tau_{th}$ according to energy conservation law (see Eqs.
(19-21)) and it leads to nonmonotonous time dependence of $E_e$
when $\gamma = 1, 10$ (this effect is absent for $\gamma =100$
when $\tau_{leak} \simeq \tau_{th}$ - see Fig. 1).

For self-consistency of found $\tau_{th}$ the radius of initial
hot spot ($\xi$ in our model) should coincide or be smaller than
diffusion length of hot electrons $\sim 2\sqrt{D\tau_{th}}$. To
make such a comparison one should know $\tau_0$, $\gamma$ and
$\alpha_{e-e}$ for NbN. Theoretical estimation with help of Eq.
(6) and $g=1$ gives $\tau_0 \simeq 925 ps$. $\tau_0$ could be also
found if one knows $\tau_{e-ph}(T_c)$ via relation $\tau_0=14
\zeta(3) \tau_{e-ph}(T_c)$ \cite{Eliashberg,Schmid2}. In thin NbN
film $\tau_{e-ph}(T_c=10 K)\simeq 16 ps$ \cite{Semenov_tau_0_NbN}
which gives us $\tau_0 \simeq 270 ps$. With last value for
$\tau_0$ and $R_{\Box}=500 Ohm$ we find $\alpha_{e-e} \simeq 1.8$.
Such a small $\alpha_{e-e}$ means (see Fig. 2(b)) that at least at
the initial stage of hot spot formation e-e scattering plays small
role and downconversion cascade and thermalization occurs mainly
via electron-phonon and phonon-electron scattering. Our estimation
of $\gamma =9$ for NbN is based on $N_{ion}= 4.8\cdot 10^{22}
cm^{-3}$ calculated from molar mass $106.9 \, g/mol$ and density
$\rho= 8.47 \, g/cm^3$ taken from Wikipedia). Therefore for NbN
$\tau_{th} \simeq 2.1 \cdot 10^{-3} \tau_0 \simeq 0.57 ps >
\tau_{|\Delta|} \simeq 0.42 ps$ and radius of initial hot spot
$2\sqrt{D\tau_{th}} \simeq 11 nm$ more than two times larger than
$\xi=4.8 nm$. These calculations show that one cannot expect
complete thermalization of electrons and phonons in NbN by time
when radius of hot spot becomes about $\xi$ and $E_{photon}=1.3
eV$.

\begin{figure}[hbtp]
\includegraphics[width=0.53\textwidth]{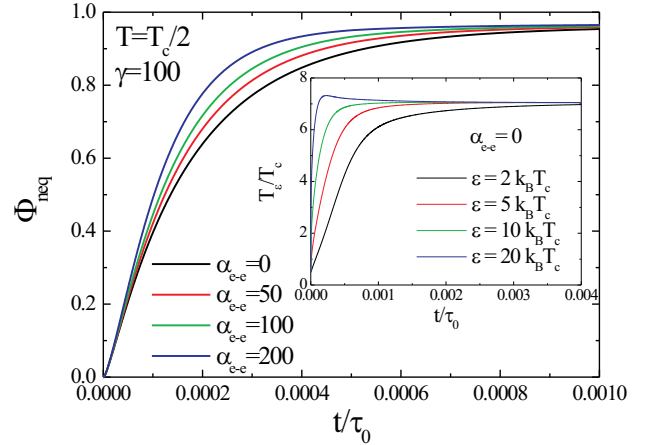}
\caption{(a) Time evolution of $\Phi_{neq}$ at $\gamma=100$ and
different $\alpha_{e-e}$. All results are obtained for phonon
bubble initial condition, $T=T_c/2$ and parameters of WSi taken
from \cite{Sidorova_arxiv}. In inset we present time evolution of
$T_{\epsilon}=\epsilon/k_Bln(1/n_{\epsilon}-1)$ at different
energies and $\alpha_{e-e}=0$. After injection of energy to the
phonon system $n_{\epsilon}$ deviates from equilibrium which leads
to different $T_{\epsilon}$. $T_{\epsilon}=T_e\simeq 7T_c$ at all
energies at $t\gg \tau_{th}\simeq 1.9 \cdot 10^{-4} \tau_0$.}
\end{figure}

We also do calculations for WSi material which demonstrates good
ability to detect single photons in optical and near-infrared
range \cite{Baek_APL}. We take parameters of WSi from
\cite{Sidorova_arxiv} ($T_c=3.4 K$, $N(0)=26.5 eV/nm^3$, $\hbar
\omega_D=34 meV$, $D=0.58 cm^2/s$, $d=3.4 nm$). Results are shown
in Fig. 3 where we use only phonon bubble initial condition and
$\gamma=100$ which is close to $\gamma=89$ expected for WSi
($C_e/C_{ph}|_{T_c}=5.65$ is calculated in Ref.
\cite{Sidorova_arxiv} with help of molar mass and density of WSi).
In this material $\tau_0 \simeq 10 ns$ if one uses theoretical
estimation from \cite{Sidorova_arxiv} or $\tau_0 \simeq 1.9 ns$ as
it follows from recent experiment where $\tau_{e-ph}$ has been
extracted from temperature dependence of magnetoconductivity
\cite{PRB_Zhang}. We adopt last value and with $R_{\Box}=595 Ohm$
it gives us $\alpha_{e-e} \simeq 5$. For $E_{photon}=1.3 eV$ we
have $\tau_{th}\simeq 1.9 \cdot 10^{-4} \tau_0 \simeq 0.36 ps \ll
\tau_{|\Delta|} \sim 1.3 ps$ for WSi and radius of hot spot
$2\sqrt{D\tau_{th}}\simeq 9.1 nm$ is close to coherence length in
WSi $\xi \simeq 8.3 nm$ which we use at calculation of $V_{init}$.

It seems that accounting of diffusion of nonequilibrium electrons
at initial stage of hot spot formation  (at $t \lesssim
\tau_{|\Delta|}$) should only decrease $\tau_{th}$. Indeed, volume
of hot spot $V\ll V_{init}$ at times $t \ll \tau_{|\Delta|}$ which
leads to higher energy concentration and, hence, faster
thermalization.

It is obvious that $\tau_{th}$ depends on energy of the photon and
$V_{init}$. This thermalization time could be estimated without
solution of kinetic equations with help of energy conservation law
(Eq. (19)) and if one associates $\tau_{th}$ with
$\tau_{e-ph}(T_e)$ for electrons having energy $\epsilon \ll
k_BT_e$. Assuming that electron and phonon distribution functions
are described by Fermi-Dirac and Bose-Einstein expressions with
$T=T_e=T_{ph}$ from Eq. (20,21) one finds Eqs. (24,25) of section
IV. For parameters of NbN and WSi materials from Eqs. (19,24,25)
it follows that absorption of photon with energy 1.3 eV in chosen
volume $V_{init}=\pi \xi^2d$ heats locally electrons and phonons
up to temperature $T_e=T_{ph} \simeq 3 T_c$ and $T_e=T_{ph} \simeq
7 T_c$, respectively, which is close to results of numerical
calculations (see insets in Figs. 2b, 3). $\tau_{e-ph}(T_e)$ could
be expressed via $\tau_0$ as $\tau_{e-ph}(T_e)=\tau_0/(14
\zeta(3))(T_c/T_e)^3$ \cite{Eliashberg,Schmid2} which gives
$\tau_{th} = \tau_{e-ph}(T_e) \simeq 2.2 \cdot 10^{-3} \tau_0$ and
$1.8 \cdot 10^{-4} \tau_0$ for these materials, which is again
close to numerical results.

We did the same calculations for normal metal (we put
$|\Delta|=0$, which in the experiment could be done by application
of relatively large magnetic field) and found very similar
results. This is not surprising, because at times $t \ll
\tau_{leak}$ deviation from equilibrium occurs at the energies
much larger than $|\Delta|$, while at $t > \tau_{th}$ main
contribution to collision integrals comes from energies $\epsilon
\simeq k_BT_e \gg |\Delta|\simeq 1.76 k_BT_c $ (see insets in
Figs. 2, 3) where spectral functions $N_1$ and $R_2$ are close to
their values in the normal state. We also expect weak dependence
on the bath temperature (if it varies in range $0 \lesssim T
\lesssim T_c$) because of large injected energy, which provides
local heating of electrons and phonons up to $T_e \gg T_c$.

We have to stress that our results are valid only at times
$t\lesssim \tau_{|\Delta|}$ when the volume of the hot spot is
smaller than $V_{init}=\pi \xi^2d$. At larger times $t \gtrsim
\tau_{|\Delta|}$ shown in Figs. 1-3 results should be considered
only as a precursor for consequent dynamics of $E_e$ and
$\Phi_{neq}$. To study the evolution of hot spot at time $t
\gtrsim \tau_{|\Delta|}$ we consider two limits. In the first
limit, limit of long thermalization time $\tau_{th} \sim
\tau_{D,w}\gg \tau_{|\Delta|}$ ($\tau_{D,w} \simeq w^2/16D -
w^2/4D \simeq 12.5-50 ps$ for strip with $w=100 nm$ and $D=0.5
cm^2/s$, depending where photon is absorbed - in the center or at
the edge of the strip), it is assumed that the largest impact on
superconducting properties occurs when hot electrons reach both
edges of the strip and simultaneously they become thermalized. We
expect such a situation in superconductors with small $D \lesssim
1 cm^2/s$ and $\gamma \leq 1$ or in materials with relatively
large $\gamma >1$ and large diffusion coefficient $D \gg 1
cm^2/s$. Because usually $\tau_{D,w} \gg \tau_{|\Delta|}$ we
expect that $|\Delta|(t)$ changes with $\Phi_{neq}(t)$ instantly
and by time $t \simeq \tau_{D,w}$ one has hot belt - region with
heated electrons and phonons up to temperature $T_e=T_{ph}>T$ and
partially suppressed $|\Delta|(T_e)$ across whole width of the
strip (see Fig. 4(a)). To calculate $T_e$ and to find the critical
current of the strip with hot belt one can use energy conservation
law and this problem is considered in section IV.
\begin{figure}[hbtp]
\includegraphics[width=0.53\textwidth]{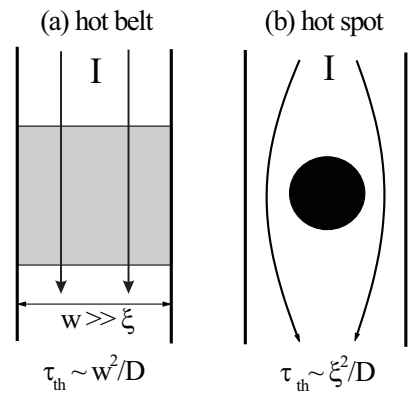}
\caption{Sketch of two situations, depending on the value of
thermalization time $\tau_{th}$ }
\end{figure}

In the second limit, limit of short thermalization time $\tau_{th}
\lesssim \tau_{|\Delta|}$, $\Phi_{neq}$ reaches maximal value
already in short time interval after energy injection (see Fig. 2)
when the radius of hot spot $R_{HS}\sim \xi \ll w$ (we expect that
it is realized in superconductors with relatively small $D
\lesssim 1 cm^2/s$ and large $\gamma \gtrsim 100$). Due to
diffusion of electrons $T_e$ decreases (and $\Phi_{neq}$ decreases
too) inside the hot spot but while $T_e>T_c$ and $R_{HS} \gg \xi$
the order parameter in the hot region is strongly suppressed. It
provides large current crowding effect around hot spot
(superconducting current avoids region with suppressed $\Delta$)
and current-carrying state may become unstable before hot
electrons reach both edges of the strip (this situation is shown
in Fig. 4(b)). We study dynamics of $\Delta$ in this limit using
modified time-dependent Ginzburg-Landau equation. We also
introduce effective temperature of electrons and phonons and solve
heat conductance equations instead of Eqs. (1,2). This limit is
studied in section V.

\section{Hot belt model}

In the hot belt model we assume that hot electrons are thermalized
among themselves and with phonons by the time when they reach
edges of the strip and form a hot belt with size $\sim w \times w$
and with local temperature $T_e=T_{ph}>T$. As a result the
critical current of the strip becomes equal to $I_c(T_e)$ because
the width and length of the belt $\sim w \gg \xi$ and proximity
effect from the regions next to the belt, where
$|\Delta|(T)>|\Delta|(T_e)$, could be neglected. We also assume
that escape time of nonequilibrium phonons to substrate
$\tau_{esc}\gg \tau_{D,w}$. The effective temperature $T_e$ may be
determined at given bath temperature $T$ and energy of the photon
from the energy conservation law

\begin{equation}
\frac{E_{photon}}{E_0w^2d}=(\mathcal{E}_e(T_e)+\mathcal{E}_{ph}(T_e))-(\mathcal{E}_e(T)+\mathcal{E}_{ph}(T))
\end{equation}
where $\mathcal{E}_{ph}(T)$ is the dimensionless energy of phonon
system per unit of volume
\begin{equation}
\mathcal{E}_{ph}(T)=\frac{1}{\gamma}\int_{0}^{\hbar\omega_D/k_BT_c}
\tilde{\epsilon}^3N_{\tilde{\epsilon}}d\tilde{\epsilon}
=\frac{1}{\gamma}\frac{\pi^4}{15}\left(\frac{T}{T_c}\right)^4,
\end{equation}
(we use that $\hbar\omega_D/k_BT_c \gg 1$ and
$N_{\tilde{\epsilon}}$ is described by Bose-Einstein function). In
Eq. (23) $\mathcal{E}_e(T)$ is the dimensionless electronic energy
per unit of volume

\begin{eqnarray}
\mathcal{E}_e(T)=\int_{|\Delta|/k_BT_c}^{\infty}\tilde{\epsilon}N_1n_{\tilde{\epsilon}}d\tilde{\epsilon}
\nonumber
\\
-\left(\frac{|\Delta|}{2k_BT_c}\right)^2\left(\frac{1}{2}+
ln\left(\frac{\Delta_0}{|\Delta|}\right)\right)=
\frac{\pi^2}{12}\left(\frac{T}{T_c}\right)^2-\mathcal{E}_s(T),
\end{eqnarray}
where $n_{\tilde{\epsilon}}$ is described by Fermi-Dirac function,
for $N_1$ we use Eq. (13) and

\begin{eqnarray}
\mathcal{E}_s(T)=\int_{0}^{|\Delta|/k_BT_c}
\tilde{\epsilon}n_{\tilde{\epsilon}}d\tilde{\epsilon}
-\int_{|\Delta|/k_BT_c}^{\infty}\tilde{\epsilon}(N_1-1)n_{\tilde{\epsilon}}d\tilde{\epsilon}
\nonumber
\\
+\left(\frac{|\Delta|}{2k_BT_c}\right)^2\left(\frac{1}{2}+
ln\left(\frac{\Delta_0}{|\Delta|}\right)\right),
\end{eqnarray}
is the gain in the energy of electrons due to their transition to
the superconducting state at $T<T_c$ ($\mathcal{E}_s(T)=0$ at
$T>T_c$).

For practical purposes one may use following interpolation
expressions for $\mathcal{E}_s(T)$ and $|\Delta|(T)$
\begin{eqnarray}
\mathcal{E}_s(T)=\left(\frac{|\Delta|(T)}{2k_BT_c}\right)^2
(1-0.053\left(\frac{|\Delta|(T)}{k_BT_c}\right)^2- \nonumber
\\
0.1\left(\frac{|\Delta|(T)}{\Delta_0}\right)^4-0.236e^{-12(1-|\Delta|(T)/\Delta_0)^{0.7}}),
\end{eqnarray}

\begin{equation}
|\Delta|(T)=1.76k_BT_c \tanh(1.74\sqrt{T_c/T-1}),
\end{equation}
which satisfy Eqs. (9, 10, 26) with accuracy better than $2 \%$.
Note, that maximal value of $\mathcal{E}_s$ is reached at about
$T=T_c/2$, where $\mathcal{E}_s^{max}\simeq 1/2$.

In presence of the superconducting current $\mathcal{E}_s$
decreases with maximal change at $I=I_{dep}(T)$. This effect could
be taken into account only numerically and it leads to small
quantitative difference to the results present in Figs. 5, 6. Here
we neglect it.
\begin{figure}[hbtp]
\includegraphics[width=0.53\textwidth]{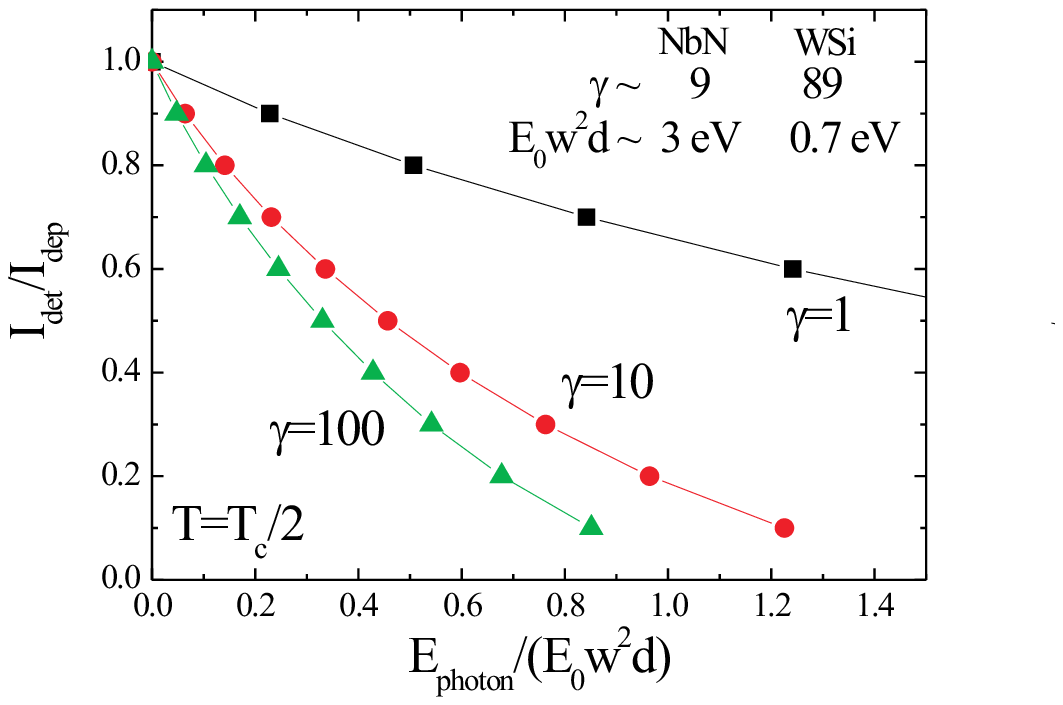}
\caption{Dependence of the detection current on the photon's
energy at different $\gamma$ and bath temperature $T=T_c/2$
calculated in hot belt model with help of Eqs. (23-29). In the
inset we present expected $\gamma$ and $E_0w^2d$ for NbN and WSi
based detectors (with $d=4 nm$, $w=100 nm$ and $d=3.4 nm$, $w=150
nm$, respectively).}
\end{figure}

To calculate at which threshold (we call it detection $I_{det}$)
current the photon drives the superconducting strip to the
resistive state one needs to know the temperature dependent
critical current, which is equal to $I_{dep}(T_e)$ in our model
and for simplicity we adopt Bardeen expression
\begin{equation}
I_{det}=I_{dep}(T_e)=I_{dep}(0)\left(1-\left(\frac{T_e}{T_c}\right)^2\right)^{3/2}.
\end{equation}

With Eqs. (23-29) it is easy to find how detection current changes
with photon's energy at given $T$ and $\gamma$. Examples of these
dependencies are shown in Fig. 5. With increasing $\gamma$
detection current drastically decreases and at large $\gamma$
practically does not depend on it, because almost all energy of
the photon goes to electronic system when $\gamma \gg 1$. At
$\gamma \lesssim 1$ only small fraction of photon's energy goes to
the electronic system and detection current is about $I_{dep}$ for
considered photon's energies.
\begin{figure}[hbtp]
\includegraphics[width=0.53\textwidth]{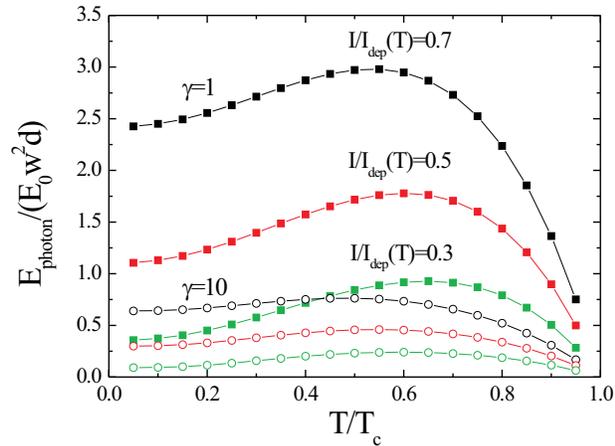}
\caption{Dependence of energy of the photon $E_{photon}$, whose
absorption drives the superconducting strip to resistive state, on
temperature for $I/I_{dep}(T)=$ 0.3, 0.5, 0.7 and $\gamma=$1, 10.
Calculations are made in framework of hot belt model.}
\end{figure}

If we fix the ratio $I/I_{dep}(T)$ the energy of the photon
$E_{photon}$, whose absorption drives the superconducting strip to
the resistive state, changes nonmonotonically with temperature
(see Fig. 6). The nature of the effect could be easily understood
with help of Fig. 7. In this figure we present $I_{dep}(T)$ (solid
curve) and show how much one should increase the temperature (by
$\delta T$) in the hot belt to transfer the strip to the resistive
state at three different temperatures and $I/I_{dep}(T)=0.5$.
$\delta T$ decreases with increase of $T$ but due to nonlinear
temperature dependence $\mathcal{E}_e(T)$ and
$\mathcal{E}_{ph}(T)$ to heat the strip from $T=0.5T_c$ up to 0.74
$T_c$ takes more energy than from $T=0$ up to $T=0.61 T_c$.
Because $\delta T \to 0$ as $T\to T_c$ there is a local maxima in
dependence $E_{photon}(T)$ (position of maxima depends on $\gamma$
and $I/I_{dep}$ - see Fig. 6).
\begin{figure}[hbtp]
\includegraphics[width=0.53\textwidth]{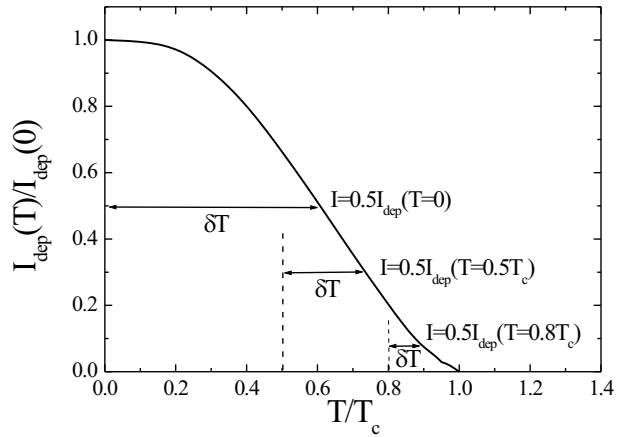}
\caption{Figure illustrates how much one should increase the
temperature in the hot belt region (at three different bath
temperatures $T=0$, $0.5T_c$, $0.8 T_c$ and three currents
$I=I_{dep}(T)/2$) to drive the superconducting strip to the
resistive state.}
\end{figure}

In above consideration we assume that at $I>I_{det}$ expanding
normal domain appears in the superconducting strip, which leads to
relatively large voltage signal in contemporary SNSPD. It is well
known that below some retrapping current $I_r(T)$ the normal
domain cannot expand and shrinks in the current carrying strip
\cite{Tinkham_PRB,Hazra}. Therefore, to see the voltage signal in
existed SNSPD the current in the strip should, at least, exceed
$I_r$. It means that for relatively large photon's energies, when
formal $I_{det}$, calculated from above equations, becomes smaller
than $I_r(T)$ real $I_{det} \gtrsim I_r$ and it should not depend
on $E_{photon}$. Support of this idea could be found in Ref.
\cite{Marsili} were WSi-based SNSPD was studied. In Fig. 3(a) of
that work dependence of photon detection efficiency (DE) on
current is present at different temperatures. One can see that in
wide temperature interval DE starts to increase at the current
which weakly depends on the temperature. This resembles weak
temperature dependence of $I_r$ at relatively low temperatures
(see for example Fig. 7(b) in \cite{Hazra}) in contrast with
noticeable temperature dependence of the critical (switching)
current in the same temperature interval (compare Fig. 3(a) of
Ref. \cite{Marsili} with Fig. 7(b) of Ref. \cite{Hazra}).

Due to the same reason one should treat results shown in Fig. 6
carefully at temperatures close to $T_c$. As $T\to T_c$ retrapping
current approaches depairing (critical) current and at some
temperature $I_r=I_{dep}$ (in Ref. \cite{Hazra} it occurs at $T
\simeq 0.82T_c$). Therefore, results present in Fig. 6 are valid
while current is larger than $I_r(T)$.

Let's make estimations for NbN and WSi. Product $E_0w^2d \simeq 3
eV$ and $\gamma \sim 9$ in case of NbN strip with $w=100$ nm and
$d=4$ nm. For WSi strip $E_0w^2d \simeq 0.7 eV$ and $\gamma \sim
89$ ($w=150$ nm, $d=3.4$ nm). According to Fig. 5 at current
$I=I_{dep}/2$ and temperature $T=T_c/2$ NbN strip would be able to
detect single photons with energy 1.35 eV, while WSi strip can
detect photons with much smaller energy 0.23 eV, if one believes
that hot belt model is valid for these materials and $\tau_{esc}
\gg \tau_{D,w}$.

\section{Hot spot two-temperature model}

In this section we study limiting case of short thermalization
time $\tau_{th} \simeq \tau_{|\Delta|}$ at the initial stage of
hot spot formation. Due to diffusion of hot electrons (one may
neglect diffusion of hot phonons due to their much lower group
velocity in comparison with one for electrons) concentration of
absorbed photon's energy in the hot spot with radius $R_{HS}>d$
decreases as $1/R_{HS}^2$. But because $\mathcal{E}_e \sim T_e^2$
(see Eq. (25)) and $\mathcal{E}_{ph}\sim T_{ph}^4$ (see Eq. (24))
the temperature of electrons drops weaker than $\sim 1/R_{HS}$.
One also should have in mind that with decreasing of $T_e$ and
$T_{ph}$ the major part of photon's energy goes to the electronic
system  - due to faster decrease of $\mathcal{E}_{ph}$ in
comparison with $\mathcal{E}_e$.

From another side diffusion time $\tau_{D}\sim R_{HS}^2/4D$
increases and one can expect that during diffusion the
nonequilibrium electrons have time for their thermalization and
electron distribution function could be described by the
Fermi-Dirac function with effective temperature $T_e \neq T$.
Indeed, when $T_e\simeq T_c$ diffusion of hot electrons is impeded
due to large $|\Delta|\simeq \Delta_0 \simeq 1.76 k_BT_c$ outside
the hot spot which should favor thermalization of electrons.

In NbN electrons and phonons are not thermalized on time scale
$\lesssim \xi^2/4D$ (see section 3). However difference between
diffusion time and thermalization time is not huge and one can
expect that despite absence of full thermalization $\Phi_{neq}$ is
relatively strong inside the growing hot spot to suppress
$|\Delta|$ substantially. Indirect prove of this idea comes from
the experiment with a magnetic field \cite{Vodolazov_PRB} which
validates hot spot model with strongly suppressed $\Delta$ in this
material. From quantitative point of view usage of 2T model for
NbN leads to smaller value of detection current (at fixed energy
of the photon) due to stronger suppression of $|\Delta|$ inside
the hot spot in comparison with the situation when electrons are
not fully thermalized.

In absence of phonon-phonon interaction thermalization of phonons
occurs only via electron-phonon and phonon-electron scattering.
Our study of initial stage of hot spot formation demonstrate
dependence of dynamics of $\Phi_{neq}$ only on the value of
injected energy to the phonon system but not on the way how it is
injected (via phonon bubble or phonon plateau initial conditions).
To account the energy adopted by the phonon system during
diffusion of electrons we assume that phonon distribution function
is described by the Bose-Einstein expression with phonon
temperature $T_{ph}$ which, in general, could be different from
$T_e$.

With above assumptions from Eqs. (1-4) one can derive (in a way
how it was done in Ref. \cite{Nagaev} for normal metal) equations
for dynamics of electron and phonon temperatures
\begin{eqnarray}
\frac{\partial}{\partial
t}\left(\frac{\pi^2k_B^2N(0)T_e^2}{3}-E_0\mathcal{E}_s(T_e,|\Delta|)
\right) = \nonumber
\\
= \nabla k_s \nabla
T_e-\frac{96\zeta(5)N(0)k_B^2}{\tau_0}\frac{T_e^5-T_{ph}^5}{T_c^3}+\vec{j}\vec{E},
\\
\frac{\partial T_{ph}^4}{\partial
t}=-\frac{T_{ph}^4-T^4}{\tau_{esc}}+\gamma\frac{24\zeta(5)}{\tau_0}\frac{15}{\pi^4}\frac{T_e^5-T_{ph}^5}{T_c},
\end{eqnarray}
where $k_s$ is a heat conductivity in the superconducting state
\begin{equation}
k_s=k_n\left(1-\frac{6}{\pi^2(k_BT_e)^3}\int_0^{|\Delta|}\frac{\epsilon^2
e^{\epsilon/k_BT_e}d\epsilon}{(e^{\epsilon/k_BT_e}+1)^2}\right),
\end{equation}
$k_n = 2D\pi^2k_B^2N(0)T_e/3$ is a heat conductivity in the normal
state and last term in Eq. (30) describes Joule dissipation
($\vec{j}$ is a current density and $\vec{E}$ is an electric
field). In superconducting state in Eq. (30) there are also
additional terms, proportional to the charge imbalance $Q\sim
\dot{\phi}+2e\varphi/\hbar$ and superconducting current density
$\vec{j}_s$, but their effect is small in comparison with Joule
dissipation and we skip them.

Note, that when both $|\Delta|$ and $T_e$ vary on time scale
comparable with $\tau_{|\Delta|}$, $|\Delta|$ is not determined by
$T_e$ via Eq. (10) or Eq. (27) and for $\mathcal{E}_s$ in Eq. (30)
one has to use Eq. (26) with independent time-dependent variables
$T_e(t)$ and $|\Delta|(t)$. As a physical consequence, variation
of $|\Delta|$ leads to heating or cooling of electrons
\cite{Vodolazov} depending on sign of $\partial |\Delta|/\partial
t$.

At derivation of Eqs. (30-32) we use Eqs. (13,14) for $N_1$ and
$R_2$ and we put $M=1$ in $I_{e-ph}$ and $I_{ph-e}$ which is
strictly valid when $|\Delta|=0$. In the superconducting state $M
\neq 1$ and it leads to increase of $\tau_{e-ph}$ and, hence,
process of cooling of electrons via their coupling with phonons
becomes longer. But because cooling of electrons is faster due to
diffusion (usually $\tau_{e-ph}(T_c)> \tau_{D,w}$) we do not
expect large influence of this effect, at least on time scale
$t<\tau_{D,w}$.

For normal metal with $T_{ph}=T$ Eq. (30) was obtained in Refs.
\cite{Nagaev,Kaganov} and in limit $|T_{e,ph}-T| \ll T$ above
equations with zero gradient terms coincide with Eqs. (13, 14)
present in Ref. \cite{Perrin}.

Appearance of the hot spot (region with suppressed $|\Delta|$) in
the strip leads to current redistribution. To find out current
distribution in each moment in time one has to solve equation $div
\vec{j}=0$, where current density $\vec{j}=\vec{j}_s+\vec{j}_n$
consists, in general, of superconducting $\vec{j}_s$ and normal
$\vec{j}_n$ parts. Superconducting current in the dirty (Usadel)
limit is described by following expression
\begin{eqnarray}
\vec{j}_s^{Us}=\frac{\sigma_n}{e\hbar}\vec{q}_s\int_0^{\infty}2N_2R_2(1-2n_{\epsilon})
d\epsilon \simeq \nonumber
\\
\simeq
\frac{\pi\sigma_n}{2e\hbar}|\Delta|\tanh\left(
\frac{|\Delta|}{2k_BT_e}\right)\vec{q}_s,
\end{eqnarray}
where $\sigma_n=2e^2DN(0)$ is the normal state conductivity.

The last expression in Eq. (33) is obtained in the limit of small
$|q_s|\ll q_s^{dep}$, omitting spatial derivative in Eq. (9) (in
this case $2N_2R_2=\pi \delta(\epsilon-|\Delta|)/2$, where
$\delta(x)$ is Dirac function) and when for $n_{\epsilon}$ one
uses Fermi-Dirac function. It turns out that this expression gives
good approximation for $\vec{j}_s^{Us}$ at all $|q_s|$ and we use
it in our calculations. When $k_BT_e \gg |\Delta|$ from Eq. (33)
one can derive well-known expression for $j_s$ in Ginzburg-Landau
model
\begin{equation}
\vec{j}_s^{GL}=\frac{\pi\sigma_n|\Delta|^2}{4ek_BT_c\hbar}\vec{q}_s.
\end{equation}

For normal component of the current density we adopt simplified
expression
\begin{equation}
\vec{j}_n=-\sigma_n \nabla \varphi,
\end{equation}
($\varphi$ is the electrostatic potential) which follows from more
general expression (see for example Refs.
\cite{Schmid,Watts-Tobin}) in the limit $k_BT_e \gg |\Delta|$.
Note, that the normal current density has large value (comparable
or larger than $j_s$) only in the region with suppressed
$|\Delta|$, which approves our choice.

To calculate effect of $n_{\epsilon} \neq n_{\epsilon}^{eq}$ or
$T_e\neq T$ on $|\Delta|$ we use modified time-dependent
Ginzburg-Landau equation which describes dynamics of complex order
parameter $\Delta=|\Delta|e^{i\phi}$. Eq. (10) is not convenient
for studying the situation when vortex (or vortices) appears in
the superconducting system, because strictly in the center of the
vortex $|\Delta|=0$, $|{\vec q}_s|=\infty$ and there is non zero
vorticity $\oint \nabla \phi dl=\pm 2\pi$ ($+$ for vortex and $-$
for antivortex). It is more convenient to deal with an equation
which operates with complex order parameter where vortices appear
naturally. Unfortunately, usual Ginzburg-Landau (GL) equation is
quantitatively valid only near $T_c$. Therefore we modify
coefficients at spatial derivative and at nonlinear term
($\Delta|\Delta|^2$) in GL equation to have temperature dependence
$|\Delta|(T)$ and $\xi(T)$ close to correct one at all
temperatures
\begin{eqnarray}
\frac{\pi\hbar}{8k_BT_c} \left(\frac{\partial }{\partial
t}+\frac{2ie\varphi}{\hbar} \right) \Delta= \nonumber
\\
\xi^2_{mod}\left( \nabla -i\frac{2e}{\hbar
c}A\right)^2\Delta+\left(1-\frac{T_e}{T_c}-\frac{|\Delta|^2}{\Delta_{mod}^2}\right)\Delta+
\nonumber
\\
+i\frac{(div \vec{j}_s^{Us}-div
\vec{j}_s^{GL})}{|\Delta|}\frac{\hbar
D}{\sigma_n\sqrt{2}\sqrt{1+T_e/T_c}},
\end{eqnarray}
where $\xi^2_{mod}=\pi\sqrt{2}\hbar D/(8k_BT_c\sqrt{1+T_e/T_c})$,
$\Delta_{mod}^2=(\Delta_0\tanh(1.74\sqrt{T_c/T_e-1}))^2/(1-T_e/T_c)$,
$A$ is the vector potential. When $T_e\to T_c$ coefficients
$\xi^2_{mod}$ and $\Delta_{mod}^2$ go to familiar GL coefficients
in the dirty limit. We check that Eq. (36) together with Eq. (33)
give the depairing current close to one which follows from the
dirty limit at all temperatures (the largest deviation $<5\%$
occurs at $T=0$) in contrast with Ginzburg-Landau depairing
current. Last term in right hand side of Eq. (36) provides
conservation of the superconducting current in the stationary
state with $\dot{\phi}+2e\varphi/\hbar=0$: $div \vec{j}_s^{Us}=0$.
If we do not include this term the stationary solution of Eq. (36)
leads to $div \vec{j}_s^{GL}=0$. Presence of hot electrons is
reflected in Eq. (36) via $T_e \neq T$ whose effect on $|\Delta|$
is analogical to effect of $\Phi_{neq} \neq 0$ in Eq. (10).

In the framework of the considered model the electrostatic
potential should be found from the current conservation law
\begin{equation}
div \vec{j}_n=-\sigma_n \nabla^2\varphi=-div \vec{j}_s^{Us},
\end{equation}

Eqs. (30, 31, 36, 37) are solved numerically for 2D
superconducting strip of finite width $w=20 \xi_c$ and length
$L=4w=80 \xi_c$. $\xi_c^2=\hbar D/k_BT_c \simeq
1.8\xi_{mod}^2(T_e=0)$ is a natural length scale in Usadel
equation when the energy is scaled in units of $k_BT_c$ and we
keep it for modified GL equation too. At the transverse edges we
use boundary conditions $\vec{j}_n|_n=\vec{j}_s|_n=0$ and
$\partial T_e/\partial n=0$, $\partial |\Delta|/\partial n=0$
while at the longitudinal edges: $T_e=T$, $|\Delta|=0$,
$\vec{j}_s|_n=0$, $\vec{j}_n|_n=I/wd$. The latter boundary
conditions model contact of the superconducting strip with a
normal reservoir being in equilibrium. This choice is not
explained by any physical reason but it is connected with the
simplest way 'to inject' the current to the superconducting strip
in numerical modelling.
\begin{figure}[hbtp]
\includegraphics[width=0.53\textwidth]{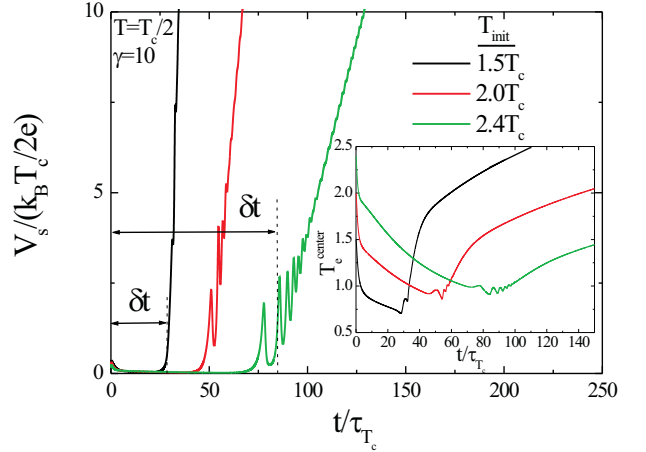}
\caption{Time dependence of voltage along superconducting strip
$V_s$ and electron temperature in the center of hot spot
$T_e^{center}$ (see the inset) calculated for different
$T_{init}$. At $t=0$ the initial hot spot appears in the center of
the strip and $I$ is slightly larger than $I_{det}$ for
corresponding $T_{init}$ (see Fig. 9). At $t\gtrsim \delta t$
there is rapid growth of the voltage because of expansion of
normal domain. Oscillations in $V_s$ and $T_e^{center}$ are
connected with nucleation and passage of vortices and antivortices
across the strip.}
\end{figure}

In numerical calculations we scale time in units
$\tau_{T_c}=\hbar/k_BT_c$ which is proportional to
$\tau_{|\Delta|}$ at low temperatures. We choose $\tau_0=900 ps
\simeq 1184 \tau_{T_c}$ which corresponds to our theoretical
estimation for NbN with $T_c=10 K$ (see discussion in section
III). We check that results change slightly with increase or
decreases of $\tau_0$ because main cooling of hot electrons is due
to their diffusion and not their coupling to phonons (which is
controlled by $\tau_0$), at least at times $t\lesssim w^2/4D$.

Based on results of section III we assume that after absorption of
the photon by the strip the hot spot with size $2\xi_c\times
2\xi_c$ appears with $T_{e,ph}=T_{init}>T$ inside the hot spot and
$T_{e,ph}=T$ outside it, while $|\Delta|=|\Delta|(T)$ everywhere.
With this initial condition at $t=0$ we study the dynamics of
$\Delta$ and $T_{e,ph}$ in the strip. In our calculations we put
$\tau_{esc}=\infty$ (effect of finite $\tau_{esc}$ is discussed in
Sec. VI).
\begin{figure}[hbtp]
\includegraphics[width=0.53\textwidth]{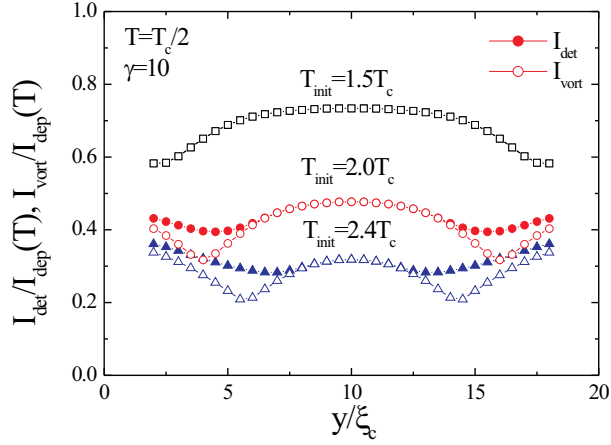}
\caption{Dependence of $I_{det}$ (at $I>I_{det}$ normal domain
expands in the superconducting strip) and $I_{vort}$ (at
$I_{vort}<I<I_{det}$ nucleation and motion of vortices does not
lead to appearance of growing normal domain) on coordinate of
initial hot spot having size $2\xi_c\times 2\xi_c$ and different
$T_{init}$. $T_{init}=1.5 T_c$ corresponds to absorption of the
photon with energy $E_{photon} \simeq 30.5 E_0\xi_c^2d \simeq 0.38
eV$, ($T_{init}=2.0 T_c \to E_{photon} \simeq 83.8 E_0\xi_c^2d
\simeq 1.04 eV$, $T_{init}=2.4 T_c \to E_{photon} \simeq 162
E_0\xi_c^2d \simeq 2.0 eV$) by NbN strip with parameters from
section III.}
\end{figure}

We found that for any $T_{init}>T$ there is a threshold current
(we call it as a detection current $I_{det}$) above which the
normal domain nucleates and expands in the strip after appearance
of the initial hot spot. Mechanism of destruction of
superconductivity depends on the position of the initial hot spot
in the strip. When it is located near the edge, at some stage of
the hot spot evolution(expansion) the vortex enters the region
with suppressed $|\Delta|$ from the nearest edge of the strip and
passes through the superconductor. After that the second, third
and so on vortices pass through the strip, the electrons are
heated because of presence of electric field $\vec{E}$ and diffuse
along the strip that leads to expansion of the resistive/normal
domain in the superconductor. When initial hot spot is located
near the center of the strip the vortex and antivortex nucleate
{\it inside} expanding hot spot, move to opposite edges of the
strip and at $I>I_{det}$ the normal domain again spreads in the
strip. Time evolution of the voltage drop along the strip and
electronic temperature in the center of hot spot located in the
center of the strip are shown in Fig. 8 for photons with different
energies (different $T_{init}$) and at current slightly above
$I_{det}(T_{init})$.

Note, that moving vortices could be nucleated in the strip with
hot spot at smaller current $I=I_{vort}<I_{det}$ but their motion
does not lead to appearance of the growing normal domain when
$I_{det} \ll I_{dep}$. Instead, after passage of one or several
vortices (number of vortices depends on the current) the
superconductivity recovers in the strip. It occurs because in the
range of the currents $I_{vort}<I<I_{det}$ cooling of hot
electrons due to their diffusion outside the moving vortex core
(where $|\vec{E}|$ is maximal) is not compensated by their heating
due to Joule dissipation $\vec{j}\vec{E}\sim I^2$.

Dependence of both $I_{vort}$ and $I_{det}$ on the coordinate of
initial hot spot and different $T_{init}$ are present in Fig. 9
(for material with $\gamma=10$). This result qualitatively
coincides with one found in quasi-stationary hot spot model (where
the present $I_{vort}$ was defined as $I_{det}$)
\cite{Zotova_SUST,Vodolazov_PRB} and resembles experimental result
from Ref. \cite{Renema_NL} (in Ref. \cite{Engel_IEEE_model}
nonmonotonous dependence $I_{det}(y)$ was predicted but without
two local minima near the edges and in that model vortices enter
the strip only via edges). Neither in Refs.
\cite{Zotova_SUST,Vodolazov_PRB} or in Ref.
\cite{Engel_IEEE_model} heating of the superconductor due to
vortex motion and condition for appearance of normal domain has
been studied.

As it is discussed in Ref. \cite{Zotova_SUST} dependence
$I_{det}(y)$ explains monotonic dependence of the detection
efficiency of SNSPD on current (when it changes from minimal
$I_{det}$ ($I_{det}^{min}$) up to its maximal value
$I_{det}^{max}$). It is interesting to note that difference
$I_{det}^{max}-I_{det}^{min}$ decreases with increase of
$T_{init}$ (energy of the photon) which resemble experimental
results found for detectors based on WSi (see Fig. 2 in
\cite{Baek_APL}), NbN (see Fig. 1 in \cite{Vodolazov_PRB}) and
MoSi (see Figs. 2,4 in \cite{Caloz_arxiv}). If one does not take
into account heating effects and associate $I_{det}$ with
$I_{vort}$ then
$I_{det}^{max}-I_{det}^{min}=I_{vort}^{max}-I_{vort}^{min}$
increases with increase of the energy of the photon (see Fig. 9
here or Fig. 5 in Ref. \cite{Zotova_SUST}).
\begin{figure}[hbtp]
\includegraphics[width=0.53\textwidth]{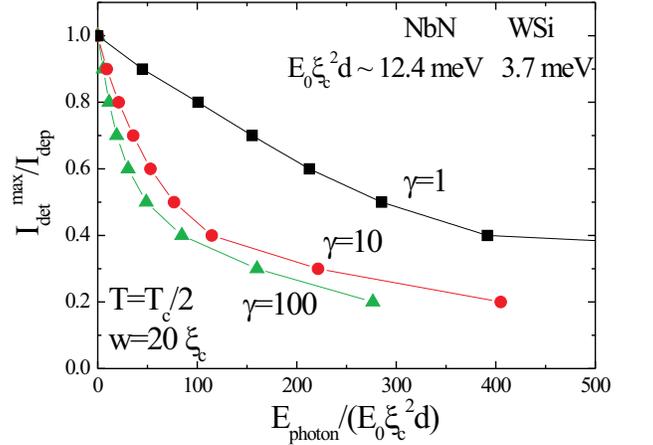}
\caption{Dependence of the maximal detection current on the
photon's energy at different $\gamma$ and bath temperature
$T=T_c/2$ calculated in 2T hot spot model. Width of the strip
$w=20 \xi_c$ ($\xi_c \simeq 6.4 nm$ for NbN with $T_c=10 K$ and
$\xi_c \simeq 11 nm$ for WSi with $T_c=3.4 K$). In the figure we
also present  value of the product $E_0 \xi_c^2d$ for NbN and WSi
based detectors with parameters from Sec. III.}
\end{figure}

When $I>I_{det}^{max}$ detection efficiency reaches maximal value
\cite{Zotova_SUST}. $I_{det}^{max}$ is located either in the
center of the strip or at its edge, depending on $T_{init}$ (see
Fig. 9 and compare it with Fig. 5 from \cite{Vodolazov_PRB} and
Fig. 4 from \cite{Zotova_SUST}). In Fig. 10 we show dependence of
$I_{det}^{max}$ on the energy of the photon. Qualitatively Fig. 10
resembles results present in Fig. 5 for hot belt model but with
one important quantitative difference. In case of superconductor
with short thermalization time (Fig. 10) one needs smaller
current, to detect the single photon or, at fixed current, the
photon with smaller energy could be detected than by the strip
with large thermalization time (Fig. 5).

\begin{figure}[hbtp]
\includegraphics[width=0.53\textwidth]{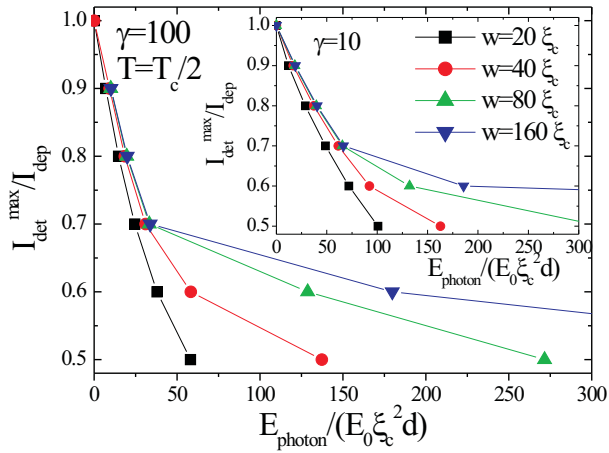}
\caption{Dependence of the maximal detection current on photon's
energy at $\gamma=100$ and $\gamma=10$ (in the inset) calculated
in 2T hot spot model for strips with different widths.
$I_{det}^{max}$ weakly depends on width of the strip when
$I_{det}^{max}/I_{dep} \gtrsim 0.7$.}
\end{figure}

In Fig. 11 we present dependence $I_{det}^{max}(E_{photon})$ for
strips with different widths. The most interesting result is that
the detection ability of the strip does not depend on its width
when $I_{det}^{max}/I_{dep} \gtrsim 0.7$. The effect originates
from current crowding around finite size spot with suppressed
$|\Delta|$ which leads to instability of superconducting state at
$I\lesssim I_{dep}$ even in the infinitely wide film
\cite{Vodolazov_PRB_1}. Due to magnetic field screening by
superconductors (in our calculations we neglect it) this effect
exists only in finite width strips with $w \lesssim
\Lambda=2\lambda_L^2/d$ where $\lambda_L$ is the London
penetration depth and screening could be neglected (for NbN film
with thickness $d=4$ nm and $\lambda_L \simeq 450$ nm, $\Lambda
\simeq 100 \mu m \simeq 15800 \xi_c$). Analytical calculations in
London model predict that static normal spot with radius $R\gg
\xi$ destroys superconducting state in the infinite film at
current $I>0.5 I_{dep}$ (see Eq. (12) in \cite{Vodolazov_PRB_1})
while calculations using stationary Ginzburg-Landau equation gives
$I \gtrsim  0.7 I_{dep}$ (see inset in Fig. 4 in Ref.
\cite{Vodolazov_SUST}). The last result is very close to our
finding where in addition we take into account expansion of the
hot spot and Joule heating.

As in case of the hot belt model we calculate how depends the
energy of the photon, whose absorption drives the strip to the
resistive state, on temperature at fixed ratio $I/I_{dep}(T)$.
From Fig. 12 one can see that this dependence is nonmonotonic one
as for the hot belt model (compare Fig. 12 with Fig. 6).
\begin{figure}[hbtp]
\includegraphics[width=0.53\textwidth]{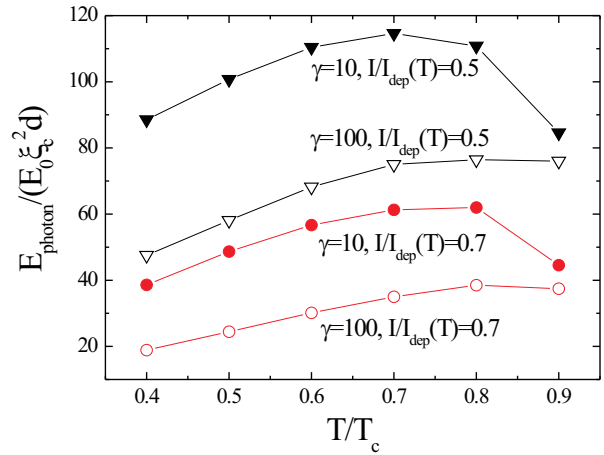}
\caption{Dependence of energy of the photon, whose absorption
drives the superconducting strip to resistive state, on
temperature at $I/I_{dep}(T)=$ 0.5, 0.7 and $\gamma=$10, 100.
Calculations are made in framework of 2T hot spot model.}
\end{figure}

And, finally, we study effect of the perpendicular magnetic field
on $I_{det}$. Qualitatively results (see Fig. 13) are similar to
theoretical findings of Ref. \cite{Vodolazov_PRB}, where $I_{det}$
is associated with $I_{vort}$ (compare Fig. 13 with Fig. 6(a) from
\cite{Vodolazov_PRB}). The only difference is that in the present
model we did not find pinning of the vortices in the strip with
$w=20 \xi_c$ at any considered $T_{init}$ (we find that in wider
strip expanding hot spot can pin vortices when $T_{init}$ is
relatively large). Small variation of $I_{det}^{min}$ in case of
large $T_{init}$ we mainly connect with weaker Joule heating as
current decreases which worsens condition for appearance of
growing normal domain. This result correlates with known effect
that retrapping current of superconducting strip has much weaker
field dependence than its critical current (see for example
current-voltage characteristics of NbN strip from Ref.
\cite{Vodolazov_SUST}).
\begin{figure}[hbtp]
\includegraphics[width=0.53\textwidth]{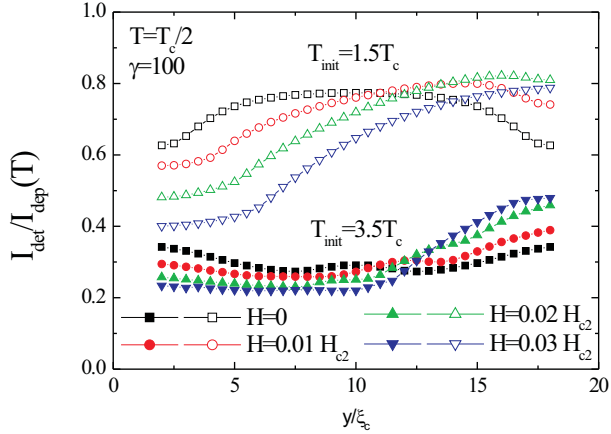}
\caption{Dependence of detection current on the position of the
initial hot spot across the strip at different magnetic fields and
two values of $T_{init}$ corresponding to absorption of photons
with different energies ($T_{init}=1.5 T_c \to E_{photon} \simeq
13.2 E_0\xi_c^2d \simeq 0.05 eV$, $T_{init}=3.5 T_c \to E_{photon}
\simeq 124.5 E_0\xi_c^2d \simeq 0.46 eV$ for parameters of WSi
strip from Sec. III).}
\end{figure}

\section{Discussion}

\subsection{Electron-phonon downconversion cascade}

Initial stage of electron-phonon downconversion cascade on time
scale of $t\lesssim \tau_{|\Delta|}$ is studied in our work using
kinetic equations for spatially uniform system with volume
$V_{init}= \pi \xi^2 d \simeq \pi D\tau_{|\Delta|}d$. In
comparison with previous works
\cite{Kresin_PRB,Kozorezov_PRB_2000} we take into account e-e
inelastic scattering and focus on the question how thermalization
time $\tau_{th}$ and distribution of photon's energy between
electronic and phonon systems depend on parameters of
superconductor and on time.

By time $t \simeq \tau_{leak}$ (which is about one picosecond for
both NbN and WSi materials and $E_{photon} \simeq 1 eV$) part of
photon's energy, initially fully absorbed by electrons, leaks to
phonons while another fraction stays with electrons. Subsequent
dynamics depends on relation between $\tau_{leak}$ and
$\tau_{th}$. When $\tau_{leak} \ll \tau_{th}$ in time interval
$\tau_{leak} < t \lesssim \tau_{th}$ both $E_e$ and $E_{ph}$
nonmonotonically vary in time (with back energy flow from phonons
to electrons and vice versus) while at $t \gg \tau_{th}$ they
become time-independent. In case of short thermalization time $
\tau_{th} \lesssim \tau_{leak}$ such a nonmonotonous dependence is
absent and at $t>\tau_{th}\sim \tau_{leak}$ both $E_e$ and
$E_{ph}$ practically do not depend on time and electrons are
thermalized.

We found both analytically and numerically that $\tau_{leak}$ is
proportional to the energy of the photon and is inversely
proportional to the square of Debye energy. Expression for
$\tau_{leak}$ (see Eq. (22)) coincides with expression for time
$\tau_1$ introduced in Ref. \cite{Kozorezov_PRB_2000} with
replacement $3E_1$ by $2\epsilon_0$ ($E_1 \gg \hbar \omega_D$ is
determined in \cite{Kozorezov_PRB_2000} as the energy at which
$\tau_{e-ph}=\tau_{e-e}$). By time $t \simeq \tau_1$ in model of
Ref. \cite{Kozorezov_PRB_2000} practically all energy of the
photon is transferred to phonon system. Our calculations show that
by time $t \simeq \tau_{leak}$ only part of photon's energy goes
to phonons and size of this part depends on parameter $\gamma$
(see Fig. 1).

Thermalization time depends on $\gamma$, strength of e-e
scattering and energy of the photon. The larger $\gamma \sim
C_e/C_{ph}|_{T_c}$ the larger part of photon's energy finally goes
to electrons and shorter $\tau_{th}$. We found that in case of
relatively large $\gamma \gtrsim 100$ and $E_{photon}\simeq 1 eV$
thermalization time may be about of leakage time even in absence
of e-e scattering. We also found that for typical low temperature
'dirty' superconducting NbN or WSi films e-e scattering plays no
role in electron-phonon energy cascade at $t \lesssim
\tau_{|\Delta|}$.


In materials with short $\tau_{th}\simeq \tau_{|\Delta|}$ the
electron-phonon downconversion cascade at $t>\tau_{|\Delta|}$ is
connected with cooling of electrons and phonons due to diffusion
of hot electrons and suppression of $|\Delta|$. We studied this
problem assuming complete thermalization of electrons at every
step of diffusion process. In this approach suppression of the
superconducting order parameter is described solely by $T_e \neq
T$ and instability of the superconducting state occurs before the
hot electrons reach the both edges of the superconducting strip.

\subsection{Effect of finite escape time and kinetic inductance}

In our calculations we neglect energy flow to the substrate, which
is controlled by escape time of nonequilibrium phonons
$\tau_{esc}$ in Eq. (2). One also should keep in mind that in
SNSPD current deviates from the superconductor and flow via the
shunt when superconducting strip/meander transits to the resistive
state. Both effects obviously should increase $I_{det}$, because
decrease of $\tau_{esc}$ enhances cooling of electrons while
decrease of current weakens Joule heating and it worsens the
condition for appearance and expansion of the normal domain. An
impression about characteristic time scales could be extracted
from Fig. 8. Normal domain expands at $t \gtrsim \delta t$ (at
that times voltage grows rapidly) and $\delta t$ increases with
decreasing $I_{det}$ due to decrease of Joule dissipation $\sim
I^2$. Therefore, when $\tau_{esc} \lesssim \delta t$ and kinetic
inductance of the detector $L_k$ is relatively small, so $L_k/R_s
\lesssim \delta t$ ($R_s$ is averaged over time interval $[0,
\delta t]$ resistance of the superconductor) these effects must be
taken into account. In Fig. 14 we show effect of finite
$\tau_{esc}$ on the energy-current relation. For large $\gamma$
escape of nonequilibrium phonons to substrate has less effect on
$I_{det}$ because of large ratio $C_e/C_{ph}|_{T_c}$, leading to
rise of time of energy transfer from electronic system to phonon
one and than to substrate.
\begin{figure}[hbtp]
\includegraphics[width=0.53\textwidth]{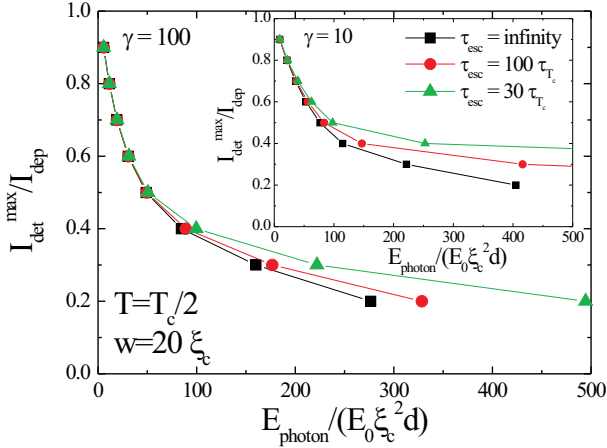}
\caption{Dependence of the maximal detection current on the
photon's energy at $\gamma=100$ and different $\tau_{esc}$
calculated in 2T hot spot model. In the inset results for
$\gamma=10$ are shown.}
\end{figure}

\subsection{Current-energy relation}

Both hot belt and 2T hot spot model predict nonlinear
current-energy relation (see Figs. 5, 10). The nonlinearity at
small energies comes from nonlinear temperature dependence of the
critical (depairing) current, energy of electron and phonon
systems and in case of hot spot model additional nonlinearity
comes from current crowding effect around hot spot in the strip
with finite width \cite{Vodolazov_PRB_RB}. Nonlinearity at high
energies originates from existence of retrapping current, below
which normal domain cannot expand in current-carrying
superconductors. Therefore at high energies $I_{det}$ should not
depend on the photon's energy (in modern SNSPD where large voltage
signal appears only when large part of superconducting strip
converts to the normal state) and it should be about of the
retrapping current of the superconducting strip. Retrapping
current goes to zero when $\tau_{esc} \to \infty$ (when length of
the superconductor is much larger than so called healing length
\cite{Skocpol_JAP}) that's why in Figs. 5, 10 there is no
saturation of $I_{det}$ at high energies (in that calculations
$\tau_{esc}= \infty$).

In Ref. \cite{Kozorezov_PRB_2015} nonlinear current-energy
relation was found (see Fig. 12 there) in the model which
resembles our hot belt model. Authors of Ref.
\cite{Kozorezov_PRB_2015} assume that electrons are thermalized
and become uniformly distributed across the strip soon after
photon absorption. The main difference between our hot belt model
and approach of \cite{Kozorezov_PRB_2015} is that we explicitly
take into account heating of phonons by absorbed photon and for
simplicity we neglect effect of the current on the electronic
energy in the superconducting state (the last effect should be
important for studying multiphoton detection
\cite{Marsili_PRB_2016} when relaxation of hot spot induced by
first photon at large times is determined mainly by
electron-phonon inelastic relaxation \cite{Kozorezov_PRB_2015} on
the background of large $|\Delta| \gg k_BT_e$). In Ref.
\cite{Kozorezov_PRB_2015} the part of the photon's energy which
goes to the electronic system is called as energy deposition
factor and it is considered as a fitting parameter which does not
depend on $E_{photon}$. From Eqs. (23-25, 29) it follows that part
of the photon's energy which goes to the electronic system does
depend on $E_{photon}$ due to nonlinear temperature dependencies
$\mathcal{E}_{e}(T)$ and $\mathcal{E}_{ph}(T)$.

Experimentally nonlinear current-energy relation was observed for
NbN, WSi \cite{Vodolazov_PRB} (in inset of Fig. 10 from
\cite{Vodolazov_PRB} results for WSi detector are extracted from
results of Ref. \cite{Baek_APL}) and MoSi \cite{Caloz_arxiv} based
detectors. To make quantitative comparison between theory and
experiment one has to know many material parameters, some of them
are known ($N(0)$, $T_c$, $R_{\Box}$, $D$) but some of them are
not ($\gamma$, $\alpha_{e-e}$, $\tau_{esc}$). The last parameters
could be extracted from additional experiments where $N_{ion}$,
$\omega_D$, $\tau_{e-e}$ and $\tau_{esc}$ could be measured.
Calculations of $\gamma$ for NbN and WSi are based on $N_{ion}$
found from molar mass and density of these materials while Debye
frequency is taken either from available experimental data (where
it varies for different phases of NbN more than in two times
\cite{SR_Zou}) or it is a result of reasonable estimation
\cite{Sidorova_arxiv}. In expression for $\alpha_{e-e}$ we put
$a=1$ (see Eq. (5)) which is not justified by any experiments or
rigorous calculations (due to their absence) for NbN and WSi.
Taking into account these circumstances and absence of reliable
value for $\tau_{esc}$ we did not make quantitative comparison
with an experiment in our work.

In theoretical paper \cite{Engel_IEEE_model} by A. Engel et. al
nearly linear current-energy relation is predicted and in Ref.
\cite{Renema_PRL} such a dependence is observed for NbN bridge in
large interval of photon's energies (in Ref. \cite{Lusche_JAP}
current-energy relations for NbN and TaN meanders look as linear
ones but they were found in narrow energy interval and could be
fitted by nonlinear functions with only one fitting parameter -
see Ref. \cite{Vodolazov_PRB_RB}). Experiment shows
\cite{Renema_APL} that for WSi bridge current-energy relation is
also nearly linear with small deviation from linear behavior at
low energies. The reason for discrepancy between experimental
results found in the meander and bridge geometry is not clear at
the moment. For example it could be connected with nonuniform
current distribution which appears naturally in the bridge those
length is comparable with its width. In such a geometry the
current density is maximal near edges of the bridge which
definitely should affect position dependent detection current and
may influence current-energy dependence quantitatively.

\subsection{Temperature-dependent cut-off wavelength}

It was found in many experiments, that detection efficiency of
SNSPD at fixed current drops very fast at wavelength larger some
critical value (it is called as cut-off wavelength $\lambda_c$ or
red boundary wavelength \cite{Engel_APL,Lusche_JAP}). Systematic
measurements of $\lambda_c(T)$ in Ref. \cite{Engel_IEEE_2013}
reveal that $\lambda_c$ decreases with increase of the temperature
(in considered temperature interval $0.05 T_c - 0.6 T_c$) when one
keeps ratio $I/I_{dep}(T)$ constant. It is in contrast with naive
expectation that as $T\to T_c$ and $|\Delta|$ decreases one needs
lower energy photon (longer wavelength) to destroy
superconductivity.

We calculate cut-off photon's energy at different temperatures and
fixed ratio $I/I_{dep}(T)$ in hot belt (see Fig. 6) and 2T hot
spot (see Fig. 12) models. Both models predict increase of cut-off
energy (decrease of $\lambda_c$) with temperature increase when
$T\lesssim T_1$ ($T_1= 0.6 T_c - 0.8 T_c$ depending on the model,
ratio $I/I_{dep}(T)$ and parameter $\gamma$). In both models the
effect mainly comes from nonlinear temperature dependence of
electronic and phonon energies (see discussion around Fig. 7).

\subsection{Photon detection at temperature near $T_c$}

As $T\to T_c$ both models predict decrease of cut-off energy
(increase of $\lambda_c$) in temperature interval $0.6 T_c-0.8 T_c
\lesssim T<T_c$ when one does not consider expansion of normal
domain. When $I<I_r(T)$ the normal domain cannot expand in the
superconducting strip and SNSPD should lose ability to detect
single photons. Because $I_r(T)=I_c(T)$ ($I_c(T)$ is the critical
current of real meander/strip) at some temperature $T^*$ close to
$T_c$ the detector cannot detect single photons at $T \gtrsim
T^*$. Actually it can stop detecting single photons even at lower
temperature. Indeed, retrapping current is determined from the
balance between Joule heating and heat removal to the substrate.
For relatively short normal domain (with length shorter than
thermal healing length $\eta$ \cite{Skocpol_JAP}) additional heat
removal comes from diffusion of hot electrons from hot spot which
increases $I_r$. From Eqs. (30,31) it follows that the healing
length at $T_e\simeq T_c$ and $|T_e-T_c| \ll T_c$ is $\eta=(2\pi^2
D\tau_0/1440\zeta(5))^{1/2}((1+\alpha)/\alpha)^{1/2}$ where
$\alpha=\pi^4 \tau_0/(450 \zeta(5) \gamma \tau_{esc})$. With used
parameters of NbN, $\tau_0=270 ps$ and $\tau_{esc}\simeq 20 ps$ we
find $\alpha \simeq 0.3$ and $\eta \simeq 29 nm \simeq 4.5 \xi_c$
which is larger or comparable to the radius of hot spot when it
drives the current-carrying superconducting strip to the resistive
state (radius of such a hot spot could be extracted for different
$E_{photon}$ from Fig. 8 - $I_{det}$ is minimal for hot spot which
touches the edge of the strip \cite{Zotova_SUST,Vodolazov_PRB}).

\subsection{Magnetic field as a probe for detection mechanism}

Current-energy relation and temperature dependence of cut-off
photon's energy following from hot belt and 2T hot spot model are
qualitatively the same. To distinguish, which model is related to
experiment one needs to make quantitative comparison, but it is
difficult to do due to lack of many material parameters. However
response of the detector to magnetic field is {\it qualitatively}
different in hot belt and hot spot models. In hot belt model (and
any model which assumes uniform, across the strip, distribution of
nonequilibrium electrons) applied magnetic field increases
detection efficiency at {\it any} current \cite{Vodolazov_PRB} (or
does not change it if DE reaches plateau at high current). In the
hot spot model due to position dependent $I_{det}$ weak magnetic
field may decrease DE in finite interval of currents
\cite{Vodolazov_PRB}. Therefore magnetic field plays the role of
qualitative probe for detection mechanism.

\subsection{Single photon detection by micron wide strip}

In the hot spot model detection of relatively high energy photon
by wide thin strip (with width up to several microns) does not
depend on its width if it can carry superconducting current larger
than $0.7 I_{dep}$ (see Fig. 11 and text around). It brings
qualitative difference with hot belt model where detection ability
has strong dependence on the width of the strip at any current.
Experimental observation of this effect could open the way for new
design of SNSPD in form of wide bridge which has much smaller
kinetic inductance in comparison with present meander-type
detectors and, hence, much shorter voltage pulses. Nowaday
detectors based on NbN or TaN have critical current up to $0.6
I_{dep}$ \cite{Lusche_JAP} which is not high enough (see Fig. 11)
for implementation of this idea.

\subsection{Single photon detection by high-$T_c$ superconducting strip}

Let us discuss perspectives of high-$T_c$ materials to be used as
active element in SNSPD. To simplify situation we use normal spot
(NS) model, neglect current crowding effect and position
dependence of detection current. In this oversimplified model the
radius of normal hot spot could be found from Eq. (23) with
replacement $w^2 \to \pi R^2$, $T_e=T_c$ and assuming that bath
temperature $T\ll T_c$ (in this case $\mathcal{E}_s(T)\simeq 0.4$)
\begin{equation}
R_{NS}=\sqrt{\frac{E_{photon}}{4\pi
dN(0)(k_BT_c)^2}}\sqrt{\frac{1}{\pi^2/12+\pi^4/(\gamma 15)+0.4}}
\end{equation}

In this model detection current linearly depends on radius of
normal spot and does not depend on its position
\begin{equation}
\frac{I_{det}}{I_{dep}}=\left(1-\frac{2R_{NS}}{w}\right)
\end{equation}
which is consequence of neglecting current crowding effect and
assumption that the photon absorbed near the edge of the strip
creates normal spot of the same shape (circle) as the photon
absorbed in the center of the strip.

This normal spot model gives order of magnitude correct estimation
when current is smaller than $0.5 I_{dep}$ (but larger than
retrapping current), $R_{NS} \gtrsim w/4$ and effect of current
crowding is relatively small (see Fig. 4 in Ref.
\cite{Vodolazov_SUST} for comparison of Eq. (39) with numerical
result where this effect is taken into account for normal spot
located in the center of strip). Indeed, for parameters of NbN
based detector from Ref. \cite{Vodolazov_PRB} which demonstrated
intrinsic detection efficiency (IDE) about unity at
$I/I_{dep}\simeq 0.5$ for photon with wavelength $\lambda = 1000
nm$ (see Fig. 2 in \cite{Vodolazov_PRB}) and $\gamma =9$ one finds
$R_{NS}\simeq 26 nm$ from Eq. (38). With help of Eq. (39) and
$w=100 nm$ it gives us $I_{det}/I_{dep} \simeq 0.5$ which is close
to experimental value. But in superconductor with $T_c =100 K$ and
the same other material parameters such a photon will create the
hot spot with radius ten times smaller ($R_{NS} \simeq 2.6 nm$)
and one will need strip with width $w \simeq 10 nm$ to detect this
photon with IDE $\simeq 1$ at $I=0.5 I_{dep}$. The actual width
should be even smaller because parameter $\gamma \sim 1/T_c^2$
(see Eq. (7)) which additionally decreases radius of normal spot.

Situation differs at currents larger than $ \sim 0.7 I_{dep}$ when
energy of the photon weakly depends on width of the strip (see
Fig. 11). For example for NbN detector from Ref.
\cite{Vodolazov_PRB} and photon with $\lambda = 1000 nm$ ratio
$E_{photon}/E_0\xi_c^2d \simeq 100$ while for superconductor with
$T_c=100 K$ and the same other parameters $E_{photon}/E_0\xi_c^2d
\simeq 10$ which means that one needs current about 0.9 $I_{dep}$
to have IDE $\simeq 1$ but with no limit for the width (while it
is smaller than Pearl length).

Above arguments show that usage of high-$T_c$ material in SNSPD
needs much narrower strip than low-$T_c$ material requires or
strip should be of very high quality to have critical current
about $90 \%$ of depairing current to detect optical or near
infrared photon with intrinsic detection efficiency about unity.

\section{Conclusion}

Our main conclusions are following:

1) After absorption of the near-infrared or optical photon by
dirty superconducting strip the thermalization time of both
electrons and phonons could be about of time variation of
magnitude of superconducting order parameter $\tau_{|\Delta|}$
when the radius of hot spot does not exceed superconducting
coherence length. Such a situation could be realized in
superconductors with relatively small diffusion coefficient $D
\simeq 0.5 cm^2/s$ and large ratio $C_e/C_{ph}|_{T_c} \gg 1$.

2) At times $t>\tau_{|\Delta|}$ the hot electrons are cooled due
to their diffusion, energy exchange with phonons and suppression
of $|\Delta|$ inside the expanding hot spot. The larger the energy
of the photon the larger is the size of hot spot where local
temperature $T_e \gtrsim T_c$ and superconducting state becomes
unstable at smaller current.

3) Instability is connected with nucleation and motion of the
vortices before the hot spot expands over whole width of the
strip. However their motion leads to appearance of the growing
normal domain only at current larger so-called detection current,
whose value depends on energy of the photon, place where photon is
absorbed, magnetic field and cannot be smaller than retrapping
current of the strip.

4) In superconductors with small ratio $C_e/C_{ph}|_{T_c} \ll 1$
detection of near-infrared or optical photon is possible only at
current close to depairing current in the strip with width $w \geq
20 \xi_c$, because only small fraction of photon's energy goes to
electrons. The same we may conclude for superconductors with large
diffusion coefficient because in this case the size of hot spot is
pretty large by time when electrons are thermalized (hot electrons
may form the hot belt across the strip) which leads to locally
smaller heating and weaker influence on superconducting
properties.

Calculations made for WSi with material parameters available from
the literature allows us to conclude that hot belt model should be
irrelevant for detectors made from this material and strip with $w
= 150 nm$ because $\tau_{D,w} \simeq w^2/16D \simeq 28 ps$ which
is much larger than $\tau_{th} \simeq 0.36 ps$ (see Section III).
The same conclusion we can make for NbN material despite absence
of complete thermalization of electrons at the initial stage of
hot spot formation. This conclusion is mainly based on
experimental results of Ref. \cite{Vodolazov_PRB} which support
the hot spot model with strongly suppressed $|\Delta|$ inside the
hot spot. Only for high energy photons, when $I_{det} \ll I_{dep}$
and size of the expanding hot spot, which drives the
superconductor to resistive state, is comparable with width of the
strip one may expect that hot belt model gives reasonable results.
The hot belt model could be also useful for study of two-photon
detection \cite{Kozorezov_PRB_2015,Marsili_PRB_2016}, when there
is time delay between absorption of two photons and hot region can
expand over whole width of the superconducting strip.

\begin{acknowledgments}

The study is supported by the Russian Foundation for Basic
Research (grant No 15-42-02365). D.Yu.V. acknowledges fruitful
discussions with Alexander Semenov and Alexander Kozorezov while
doing this work.

\end{acknowledgments}


\end{document}